\providecommand{\reserveinserts}[1]{}
\documentclass[VANCOUVER,LATO2COL]{WileyNJDv5}

\articletype{Invited Review}%

\received{Date Month Year}
\revised{Date Month Year}
\accepted{Date Month Year}
\journal{Journal}
\volume{00}
\copyyear{2026}
\startpage{1}

\usepackage{xcolor}
\usepackage{xstring}
\usepackage{etoolbox}
\definecolor{darkcopper}{HTML}{7B4B2A}
\definecolor{lightcopper}{HTML}{C77B58}
\usepackage{textcase}

\makeatletter
\renewcommand{\filtername}[1]{\def\temp@@name{#1}\futurelet\next\check@space}
\def\filtername#1\@nil{#1}
\makeatother

\begin{document}

\title{From London to Morse via Binnig, Quate, and Gerber}
% \title{Wiley New Journal Design version 5 (NJD-v5): As of Feb. 12, 2024, you must set the compiler as XeLaTex or XeLaTex for compatibility with all the fonts; set Tex Live version as 2022; Tex Live version 2023 won't work}

\author[1]{Sofia~Alonso~Perez}
\author[1]{Matthew~O.~Blunt}
\author[1]{Frederick~Carlisle}
\author[2]{Neil~R.~Champness}
\author[1]{Janette~L.~Dunn}
\author[1]{Matthew~Edmondson}
\author[1]{Rowan~Evers}
\author[1]{Connor~Fields}
\author[1]{Subhashis~Gangopadhyay}
\author[1]{James~Hayton}
\author[1]{Samuel~P.~Jarvis}
\author[1]{Filipe~Junqueira}
\author[3]{Lev~Kantorovich}
\author[1]{Brian~Kiraly}
\author[4]{Natalio~Krasnogor}
\author[1]{Ioannis~Lekkas}
\author[1]{Morten~M\o ller}
\author[1]{Philip~Moriarty}
\author[5]{Chris~Pakes}
\author[1]{Emmanuelle~Pauliac-Vaujour}
\author[1]{Oliver~Phillips}
\author[6]{Adrian~Radocea}
\author[1]{Philipp~Rahe}
\author[1]{Mohammad~Abdur~Rashid}
\author[7]{Hongqian~Sang}
\author[1]{Alex~Saywell}
\author[1]{Nikhil~Seeja~Sivakumar}
\author[1]{Peter~Sharp}
\author[1]{Andrew~Stannard}
\author[1]{Julian~Stirling}
\author[1]{Adam~Sweetman}
\author[1]{Simon~Taylor}
\author[1]{Richard~A.J.~Woolley}
\authormark{MORIARTY \textsc{et al.}}
\titlemark{From London to Morse, via Binnig, Quate and Gerber}
\address[1]{\orgdiv{School of Physics and Astronomy}, \orgname{University of Nottingham (UoN)}, \orgaddress{\state{NG7 2RD}, \country{UK}}}
\address[2]{\orgdiv{School of Chemistry}, \orgname{University of Nottingham}, \orgaddress{\state{Nottingham NG7 2RD}, \country{UK}}}
\address[3]{\orgdiv{Department of Physics}, \orgname{King's College London}, \orgaddress{\state{London}, \country{UK}}}
\address[4]{\orgdiv{School of Computer Science}, \orgname{University of Nottingham}, \orgaddress{\state{Nottingham NG7 2RD}, \country{UK}}}
\address[5]{\orgdiv{Department of Mathematical and Physical Sciences},\orgname{
La Trobe University}, \orgaddress{\state{Bundoora 3086, VIC,}\country {Australia}}}
\address[6]{\orgdiv{Department of Materials Science and Engineering},\orgname{Cornell University}, \orgaddress{\state{Ithaca, New York 14853},\country{USA}}}
\address[7]{\orgdiv{School of Physics and Technology, Centre for Electron Microscopy and MOE Key Laboratory of Artificial Micro- and Nano-structures}, \orgname{Wuhan University}, \orgaddress{\state{Wuhan, 430072}, \country{China}}}

\corres{Note that authors are listed in alphabetical order. Corresponding author: Philip~Moriarty, School of Physics and Astronomy, UoN. \email{philip.moriarty@nottingham.ac.uk}.}

\presentaddress{The addresses listed here are for the authors at the time of publication of the work reviewed in this paper. Present addresses, where known and where applicable, are given in the Acknowledgements section.}

%\fundingInfo{Text}
%\JELinfo{ejlje}

\abstract[Abstract]{In their landmark paper introducing the atomic force microscope [Phys. Rev. Lett. \textbf{56}, 930 (1986)], Binnig, Quate, and Gerber presciently anticipated that the technique would ultimately be capable of probing interactions running the gamut from weak van der Waals interactions to strong covalent bonding. They also highlighted that the tip-sample forces central to AFM are present, and often highly influential, in scanning tunnelling microscopy; indeed, this realisation directly inspired the invention of the force microscope. In this perspective for the \textit{Forty Years of AFM} special issue, we review selected aspects of two decades of work from our group at the University of Nottingham that span the force range highlighted by BQG and are united by a common, central theme: the probe as active participant rather than passive observer. Our selection of results also covers length- and correlation-scales from the microscopic right down to the single chemical bond limit, tracking a spectrum of interactions from van der Waals/Hamaker forces, through hydrogen bonding, to covalent bonds and, finally, atom-by-atom assembly of metal clusters via vertical tip-sample transfer. Echoing BQG's own observations on the prevalence of probe-sample forces in STM, we also discuss recent evidence that tip-induced heterogeneity underpins first-passage dynamics in molecular diffusion and highlight the challenges in acquiring non-invasive measurements of diffusion barriers for adsorbed molecules that are readily perturbed by the probe. We close with a perspective on machine learning's growing role in automating tip-driven atomic and molecular manipulation.}


\maketitle

\renewcommand\thefootnote{}
\footnotetext{\textbf{Abbreviations:} SPM, scanning probe microscopy; AFM, atomic force microscopy; STM, scanning tunnelling microscopy; DFT, density functional theory}
\renewcommand\thefootnote{\fnsymbol{footnote}}
\setcounter{footnote}{1}

\section{40 years after: Revisiting BQG}\label{sec1}
In the closing paragraphs of their groundbreaking 1986 paper\cite{Binnig1986}, Binnig, Quate, and Gerber (BGQ) looked beyond the immediate demonstration of the capabilities of the nascent atomic force microscope and considered the range of tip-sample forces that it would one day measure: ``\textit{The interatomic forces therefore range from $10^{-7}$~N for ionic bonds to $10^{-11}$~N~for van der Waals bonds and down to perhaps $10^{-12}$~N for some of the weaker forces of surface reconstruction. The limiting sensitivity of our instrument is far less than these values. Therefore, we should be able to measure all of the important forces that exist between the sample and adatoms on the stylus.}'' This is a remarkably farsighted appraisal of the potential of the AFM. The tip-sample forces that BQG estimate here -- and, of course, the invention of the AFM instrument itself -- represent the foundation on which the entire force microscopy field was built. Within a year, Martin, Williams and Wickramasinghe~\cite{Martin1987} drew directly on BGQ's paper to introduce the non-contact atomic force microscope, which further widened the scope, capabilities, and potential applications of the instrument. It was with this dynamic mode of operation that our group's own AFM-based research began in the early 2000s --  a departure, at the time, from a background rooted in ultra-high~vacuum (UHV) scanning tunnelling microscopy (and synchrotron-based spectroscopies).

The selection of work that follows spans the force range highlighted by BQG. We begin at the weakest end of the scale — with dispersion forces acting collectively across a two-dimensional nanoparticle assembly, sufficient together to reshape a surface but individually much softer than any covalent bond. From there we move progressively down in length- and up in energy-scale, probing dispersion interactions between two C$_{60}$ molecules, hydrogen bonding, the making and breaking of single covalent bonds, and culminating in atom transfer from the tip apex. Throughout, the same question recurs in different guises: what role does the probe itself play in determining just what an SPM microscopist observes and measures? Looking forward -- perhaps not quite to the next forty years, but to an horizon of a decade or so -- we ask a related question: to what extent can AI complement user intelligence in gaining greater control of the imaging, spectroscopic, and manipulation capabilities of the probe?

\section{Nanomechanical coarsening}\label{sec2}

Our first AFM experiments were carried out not under vacuum but in air\cite{Moriarty2002,OShea2002, Martin2004, Hayton2007} -- and, in some cases, under fluid -- using tapping-mode operation to image and, subsequently, to actively drive the evolution of nanoparticle assemblies held together by London dispersion forces. An example is shown in Figure 1 -- tapping mode images of a one-micron-square region of a Au-nanoparticle submonolayer film, deposited from a nonane solution onto a native-oxide terminated Si(111) substrate\cite{Blunt2007}. (The Au nanoparticles, of $\sim$ 2 nm diameter, are octanethiol-passivated, and were synthesised using the method of Brust \textit{et al}\cite{Brust1994}.) Due to the volatility of the solvent, it dewets rapidly, driving the thin nanoparticle-solvent film far from equilibrium and into the spinodal regime of the phase diagram\cite{Ge2000,Rabani2003, Sztrum2006, Martin2007, Stannard2008, Stannard2011}. This imprints the characteristic spatially correlated ``labyrinthine'' pattern seen throughout Fig. 1. 

\begin{figure}[t!]
\centerline{\includegraphics[width=0.45\textwidth]{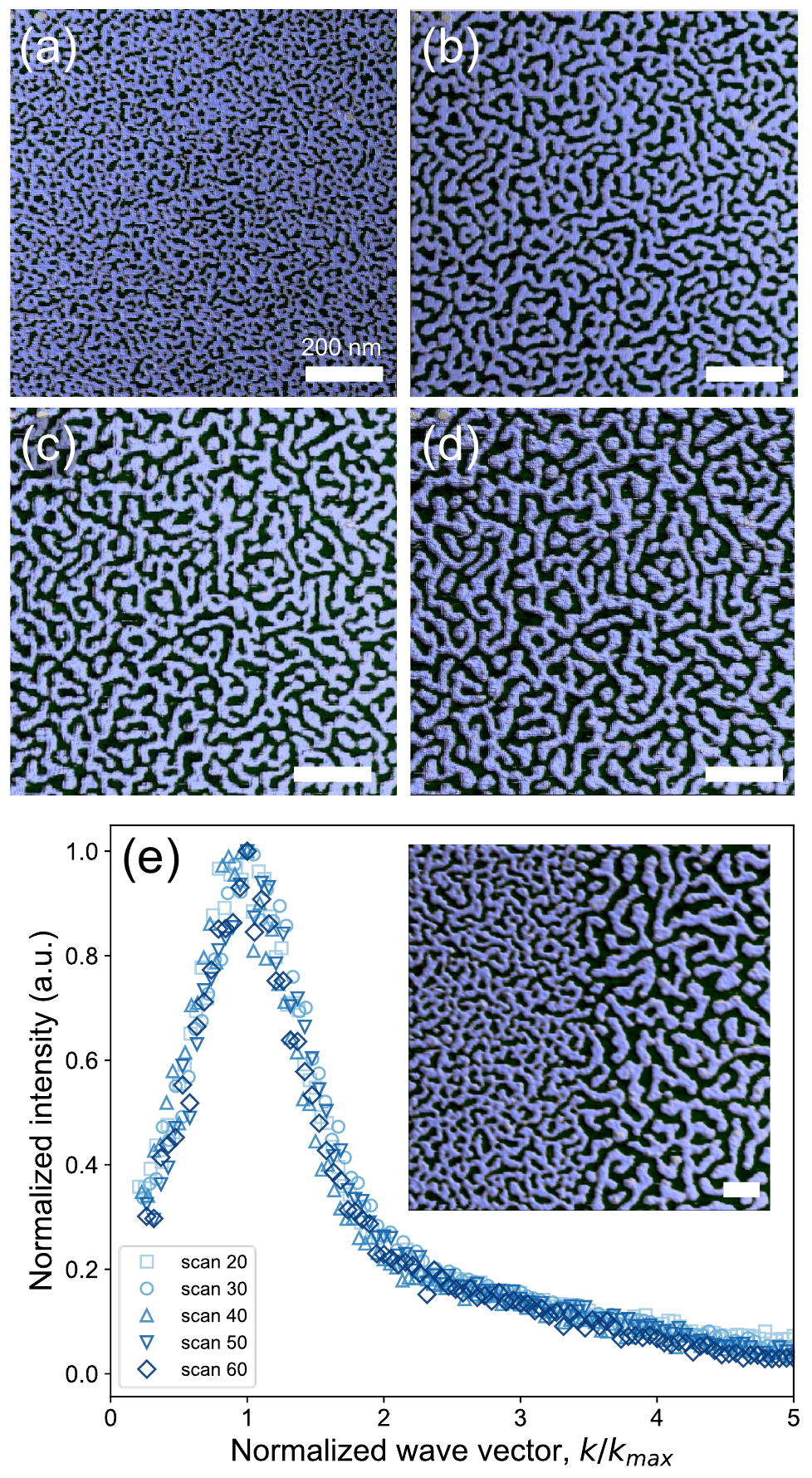}}
\caption{\textbf{Tip-driven coarsening of non-equilibrium nanostructure. (a)-(d)}. 1 $\mu\text{m}^2$ tapping-mode (TM) AFM images (taken at two-hour intervals from a continuous eight-hour scanning sequence) of the 
evolution of a 0.55~monolayer coverage of octanethiol-passivated Au nanoparticles. The interconnected labyrinthine morphology results from solvent dewetting in the spinodal limit\cite{Ge2000,Rabani2003,Sztrum2006, Martin2007, Stannard2008,Stannard2011} and, as such, is characterised by a well-defined spatial correlation length. \textbf{(e)} When rescaled to the peak wavevector, $k_{\mathrm{max}}$, radially averaged Fourier transforms of the TM-AFM images collapse onto a time-independent master curve -- the signature of self-similar coarsening. Here, however, the nanoparticle assembly is mechanically, not thermally, coarsened. \textbf{Inset:}~The delineation between the probe-coarsened region following a total of 8 hours of imaging and the surrounding
surface (scanned once, left-hand side of the image) is clear. The scale bar in each image is 200 nm.}\label{fig1}
\end{figure}

\begin{figure*}[b!]
\centerline{\includegraphics[width=0.95\textwidth]{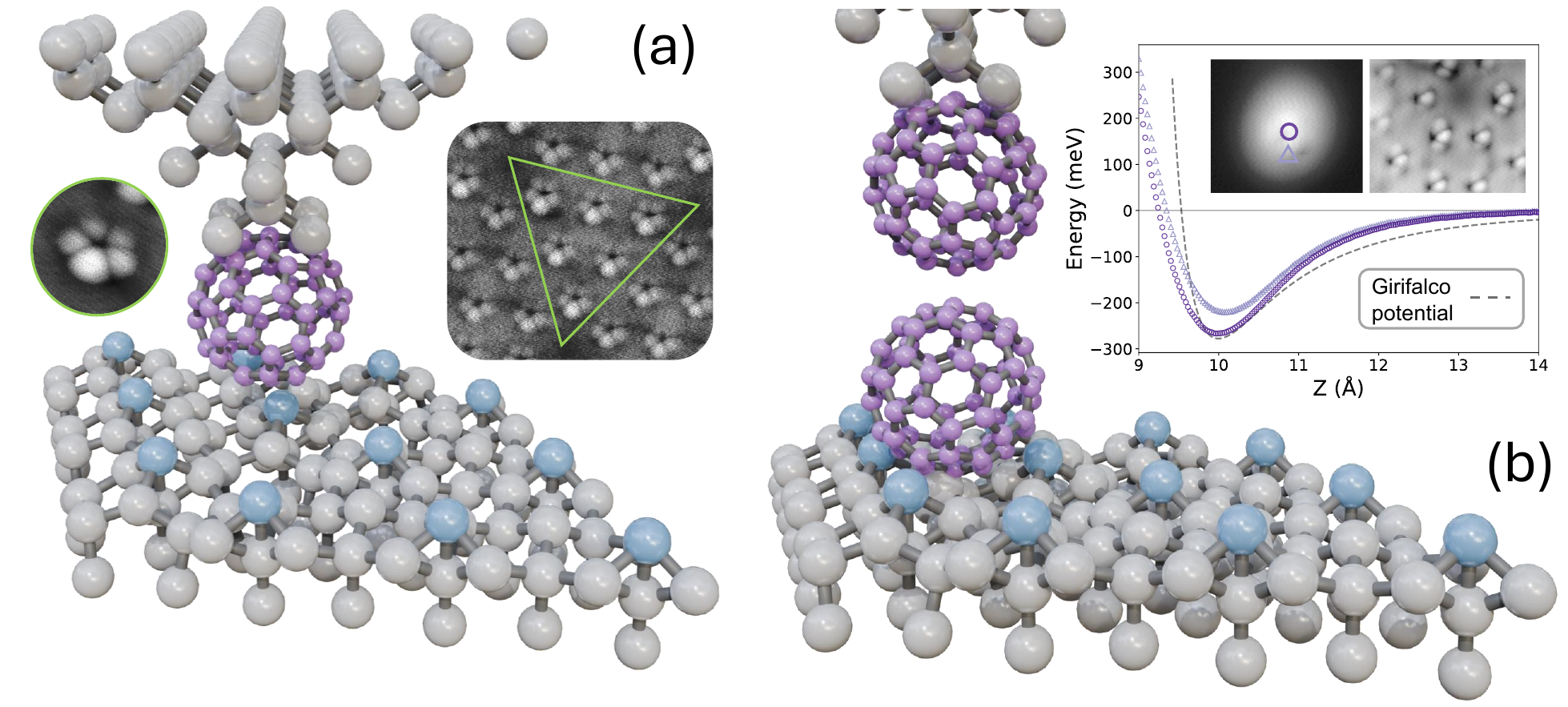}}
\caption{\textbf{Measuring dispersion energy at the single molecule level.} After transferring a C$_{60}$ molecule from the Si(111)-(7x7) surface to the qPlus AFM/STM tip\cite{Chiutu2012,Sweetman2016}, \textbf{(a)}, the adatoms of the Si(111)-(7x7) reconstruction (shaded in blue in the model) are no longer imaged as circularly symmetric maxima. Instead, each appears as a five-lobed feature (inset in circle on left). In the inset on the right of Fig. 2(a), six adatoms, comprising one half of the (7x7) unit cell, are highlighted by the green triangle.The darker patches at the vertices of the triangle are the corner-holes of the (7x7) reconstruction. \textbf{(b)} A surface-adsorbed C$_{60}$ molecule is located using the buckyball-terminated tip, enabling a qPlus AFM measurement of the pair potential (inset). The depth and overall width of the measured potential are in good agreement with the Girifalco potential, albeit with deviations in both the attractive and repulsive regimes. Measurements at different positions across the molecule (compare open circles and open triangles) show that the depth of the potential can vary significantly. From the image inset on the right of the pair potential plot the orientation of the tip-adsorbed molecule can be determined.}\label{fig2}
\end{figure*}

Figs. 1(a)-(d) are four frames, spaced two hours apart, from an AFM movie comprising a total of 62 images of the same sample area over eight hours of acquisition time. Although the overall fractional area coverage ($\sim 0.55$) remains constant throughout the entire sequence, there is a clear and dramatic change in the characteristic correlation length of the pattern. Radially-averaged Fourier transforms (Fig. 1(e)) of the images can be rescaled and collapsed onto a time-independent master curve
\begin{equation}
S(k,t)=k_{\mathrm{max}}(t)^{-2}F(k/k_{\mathrm{max}}(t)),
\end{equation}
where $k_{\mathrm{max}}$ is the position of the peak in the transform and $F(k/k_{\mathrm{max}})$ is the master curve. In other words, the morphology of the far-from-equilibrium nanoparticle assembly is characterised at a given time, i.e. given scan number, by a single length scale $L(t)$. Moreover, the time-evolution of the peak wavevector provides considerable insights into the coarsening process\cite{Blunt2007, Rabani2003,Sztrum2006}. A once-scanned area of the nanoparticle assembly alongside an adjacent region subjected to all sixty-two scans is shown in the inset to Fig. 1(e).

London dispersion forces\cite{London1930} are at play at a number of different levels in Fig 1. They hold the nanoparticle assembly together, are responsible for its adsorption on the native-oxide-terminated silicon substrate, and underpin its interaction with the AFM tip. A natural question is whether the probe-mediated mechanical coarsening we see is driven by vertical tip-sample transfer of nanoparticles, or by lateral, tip-driven diffusion — via local heating or direct dragging. Local heating is hard to sustain as an explanation of the dominant mechanism: contact during each tap of our 70~kHz, 2~N/m cantilever is extremely brief, making significant thermal activation over the local London dispersion-derived interparticle binding energy implausible. Our data instead point to transfer via direct mechanical tip-sample interactions: the seamless boundary between scanned and unscanned regions (inset to Fig.~1e), conservation of total nanoparticle mass, and a coarsening exponent of $1/2$, characteristic of Ostwald ripening limited by interface kinetics\cite{ostwald1901,Rabani2003,Sztrum2006,Blunt2007}, all indicate that attachment and detachment at island edges, not lateral diffusion, is rate-limiting. Unlike traditional thermal coarsening, the tip acts as an exceptionally local energy source (rather than a globally applied heat bath), allowing the correlation length of the assembly to be ``dialled in'' directly, simply by choosing how long, and where, the probe scans.

\section{Dispersion vs repulsion: C$_{60}$-C$_{60}$}\label{sec2}
The nanoparticle assembly of the previous section is essentially a microscopic many-body system whose coarsening was governed by the collective, statistical effect of integrated dispersion forces. A natural next step is to isolate this type of interaction at its simplest: a single pair potential for two molecules subject to London dispersion\cite{London1930, Girifalco1992}. We achieved this by transferring a single C$_{60}$ molecule from a Si(111)-(7$\times$7) surface onto the apex of a qPlus\cite{Giessibl1998, Giessibl2011, Giessibl2019,qplussensor} AFM tip (Fig. 2(a))\cite{Chiutu2012,Sweetman2016}. A second buckminsterfullerene molecule adsorbed on the surface below was subsequently located and the fullerene-terminated tip used to measure the C$_{60}$-C$_{60}$ pair potential (Fig. 2(b)), in ultrahigh vacuum at 4.5~K.

Inset to Fig. 2(a) are qPlus AFM images of the Si(111)-(7x7) reconstruction after a single C$_{60}$ molecule has been transferred to the tip\cite{Chiutu2012, Sweetman2016}. The spatial extent of the dangling bond orbitals for the adatoms of the (7x7) reconstruction (shaded in blue in the ray-traced models of Fig. 2) is significantly smaller than that of the $\sim$ 1 nm diameter of the C$_{60}$ cage, and, thus, the surface images the tip and not \textit{vice versa}. This type of inverse imaging had been successfully applied in STM studies involving tip-adsorbed C$_{60}$\cite{Schull2009,Schull2011, Lakin2013}, but the qPlus sensor enables the atomic structure of the molecular face closest to the surface to be resolved, and, hence, the molecular orientation to be determined with a precision beyond that available via local density of states (LDOS) imaging alone. An image of a single adatom acquired with a C$_{60}$-terminated tip is shown in the inset on the left of Fig. 2(a); the five-lobe pattern demonstrates that the molecule is adsorbed with a pentagon facing the Si(111) surface. One of the atoms of the pentagonal face consistently appears brighter in the constant frequency shift images (see both insets in Fig. 2(a)), indicative of a stronger tip-sample interaction at that cage site. Density functional theory (DFT) calculations accounted for this and showed that the Si-C interaction underpinning the imaging process was sufficiently high to drive shifts and distortion of the C$_{60}$ cage at the end of the tip\cite{Chiutu2012}.

We then positioned the C$_{60}$-terminated tip above a fullerene molecule adsorbed on the Si(111)-(7x7) surface (Fig. 2(b)). By ramping the tip-sample separation $z$, simultaneously measuring the variation in frequency shift, $\Delta f$, and then inverting the $\Delta f(z)$ data using the well-established Sader-Jarvis method\cite{SaderJarvis, Sweetman2014b}, we extracted $U(z)$, the intermolecular potential at different submolecular positions (inset to Fig.~2(b))\cite{Chiutu2012}. Alongside the experimental potential energy curves, the semi-empirical Girifalco pair potential for C$_{60}$\cite{Girifalco1992} is plotted as a dashed line. It is important to note that the Girifalco potential plotted in the figure is not a fit -- there is no adjustment of the form or parameterisation developed by Girifalco. Instead, a rigid shift of the experimental curve along the $z$ axis is applied to align with the minimum of the Girifalco potential. The latter not only accurately describes the depth and curvature of the experimentally measured C$_{60}$-C$_{60}$ potential around the minimum but its overall rate of decay with increasing distance beyond that minimum is also broadly comparable. Deviations are to be expected given that the Girifalco potential does not take into account the molecular environment. Each molecule is adsorbed on a substrate, which is either the tip (almost certainly silicon-terminated prior to the adsorption of a fullerene molecule) or the Si(111)-(7x7) substrate. With this in mind, the relaxation seen in the repulsive part of the potential is unsurprising, as, indeed, is the deviation in the attractive dispersion regime from the Girifalco model. Covalent bonding of C$_{60}$ to silicon\cite{Cepek1999, Gangopadhyay2009} will influence its polarisability.  

\begin{figure*}[t!]
\centerline{\includegraphics[width=1\textwidth]{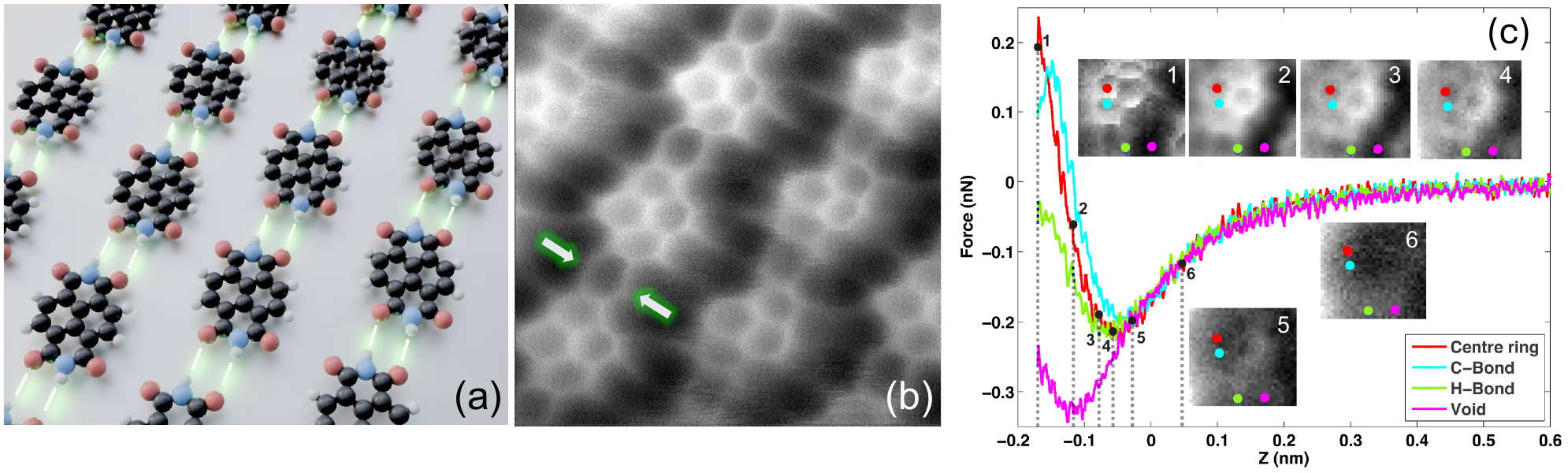}}
\caption{\textbf{H-bond or not? Intermolecular artefacts in qPlus AFM images of a hydrogen-bonded assembly.} ``Filaments'' appear between naphthalene tetracarboxylic diimide (NTCDI) molecules at the locations where hydrogen bonds (HBs) are expected\cite{Sweetman2014}. \textbf{(a)} Model of the NTCDI lattice where the hydrogen bonds between molecules are highlighted in green. (Black: carbon; grey: hydrogen; red: oxygen; blue: nitrogen). \textbf{(b)}~Constant height qPlus AFM image, acquired at 77~K, of a hydrogen-bonded NTCDI island on the Ag:Si(111)$(\sqrt{3} \times \sqrt{3})\text{R}30^{\circ}$ surface. Image size: 2.1~nm by 2.0~nm; oscillation amplitude: 275~pm. Grey scale range spans -7~Hz to -21~Hz. Green arrows highlight intermolecular features at expected positions of hydrogen bonds. \textbf{(c)} Short-range force curves recorded at 5 K over an NTCDI carbon ring centre (red), a C–C bond (cyan), a hydrogen-bond region (green), and an intermolecular void (purple), all extracted from the same atom-tracked\cite{Rahe2011} force-curve grid acquired over the course of 29~hours. Images 1–6 show constant-height slices through the grid at the tip–sample separations marked 1–6 on the force curves. Note that clear intermolecular contrast only emerges once the gradient turns negative. Data were acquired with an oscillation amplitude of 110~pm.}\label{fig3}
\end{figure*}

A natural extension of this ``single-point'' measurement is to ask how the potential varies not just with separation, but with the relative orientation of the two molecules. This was addressed by mapping the full three-dimensional potential between a C$_{60}$-terminated tip and a second, surface-adsorbed C$_{60}$ molecule with sub-\AA{}ngstrom spatial resolution\cite{Sweetman2016}. Rather than examining individual $U(z)$ curves in isolation, a method to visualise the variation in equilibrium binding energy, $U_{\mathrm{min}}$, for different molecular orientations was introduced. Changes in the relative orientation of the tip- and surface-adsorbed molecules produced shifts in $U_{\mathrm{min}}$ of order 60 meV. Decomposing this variation, via both a simple Lennard-Jones model and dispersion-corrected DFT, into its attractive and repulsive contributions produced a somewhat counter-intuitive result\cite{Sweetman2016}: the observed variation is dominated by repulsive, rather than dispersive, interactions, even at separations beyond the equilibrium value where the net force remains attractive. In other words, the orientational dependence of the C$_{60}$--C$_{60}$ potential is predominantly due to the onset of Pauli repulsion between near-facing atoms, with the underlying variation in dispersion interaction largely masked from direct measurement.

\section{``If it looks like a duck...''}
A now heavily-cited paper published in 2013 by Zhang \textit{et al.}~\cite{Zhang2013}, had the striking title ``\textit{Real-Space Identification of Intermolecular Bonding with Atomic Force Microscopy''} and generated quite some debate and controversy in the AFM community (which, to some extent, persists to this day). Following the pioneering demonstration, four years earlier, of exceptionally high-resolution \textit{intra}molecular contrast by Leo Gross and co-workers\cite{Gross2009} -- at IBM~R\"uschlikon, fittingly in the context of this retrospective -- a number of groups, including our own, were keen to explore whether \textit{inter}molecular contrast of comparable spatial resolution might also be achievable. A particularly intriguing question arose: could hydrogen bonds be resolved in real space? The authors of ``\textit{Real-Space Identification...''} were clear about where they stood: the observed bond contrast was interpreted via density functional theory calculations, and explained in terms of the ``\textit{electron density contribution from the hybridized electronic state of the hydrogen bond}''\cite{Zhang2013}. 

At the time of publication of Zhang \textit{et al.}'s paper we were writing up a manuscript, later published as Sweetman~\textit{et~al.}\cite{Sweetman2014}, that focussed on qPlus AFM imaging of the molecular system shown schematically in Fig. 3(a): a 2D self-assembled lattice of hydrogen-bonded naphthalene tetracarboxylic diimide (NTCDI) adsorbed on Ag:Si(111)$(\sqrt{3} \times \sqrt{3})\text{R}30^{\circ}$. (We chose this surface because of its passivated, low free energy character with respect to molecular adsorption (and its ease of preparation). It's a substrate we've exploited extensively over the years for studies of the self-assembly of a variety of other molecules including fullerenes, phthalocyanines, and linear acenes). Submolecular resolution (Fig. 3(b)) was attainable despite no deliberate attempt to functionalise the tip\footnote{It is worth noting that a similar strategy is also possible on the more reactive Si(111)-(7x7) surface. Two chemically distinct tip apices were found to be capable of providing submolecular contrast in that case\cite{Sweetman2014PRB}.} via, for example, CO pick-up\cite{Gross2009}. Instead, routine scanning was sufficient to yield the appropriate tip termination, which we attributed to an ``oxygen-down'' NTCDI molecule at the apex\cite{Sweetman2014}. 

It was, however, the ``filamentary'' features between the molecules, highlighted in Figs. 3(a) and 3(b), that were the focus of our study. Those appear exactly where N-H$\cdots$O hydrogen bonds are expected between NTCDI molecules. (Weaker inter-row contrast at the positions of \mbox{C-H$\cdots$O} hydrogen bonds is also observed in Fig. 3(b).) The genuinely perplexing aspect of our data was not so much the presence of contrast at the expected hydrogen-bond positions -- although that in itself was indeed somewhat surprising -- but its anomalously high intensity: the features were almost as bright as the intramolecular C–C bonds, despite the bare electron density at a hydrogen bond being, by our own calculations\cite{Sweetman2014}, over an order of magnitude weaker than that at a carbon-carbon bond. Our interpretation at the time was that the relevant quantity was not the total electron density in isolation, but the density redistributed specifically by the coupled tip-sample system as the probe approaches~\cite{Sweetman2014}. 

It turned out that this was far from the complete picture when it came to the role of the tip-sample interaction. While our paper was in the final stages of the peer review process, we spotted an abstract\cite{DPGabstract} for the upcoming 2014 Deutsche Physikalische Gesellschaft (DPG) conference (in Dresden that year) that concluded: `` \textit{Here we show, that sharp features measured in ... experiments, especially apparent intermolecular bonds, can be explained almost exclusively by simple geometrical model using pairwise potential (Lennard-Jones, Morse) when
relaxation of probe position is considered}.'' Our interest was, of course, piqued and we quickly contacted Prokop Hapala and Pavel Jelinek (Czech Academy of Sciences), the first and last authors of that abstract, respectively. That led to a great deal of discussion about the origin of the intermolecular hydrogen-bond features in our, and others', images. Shortly afterwards, Pavel and colleagues organised a lively, well-attended workshop in Prague (Feb.~2015) on high resolution AFM imaging that, like the paper that stemmed from the DPG abstract\cite{Hapala2014}, has resonated over the years and been highly influential in the field\cite{Oinonen2024}.

\begin{figure*}[b!]
\centerline{\includegraphics[width=0.85\textwidth]{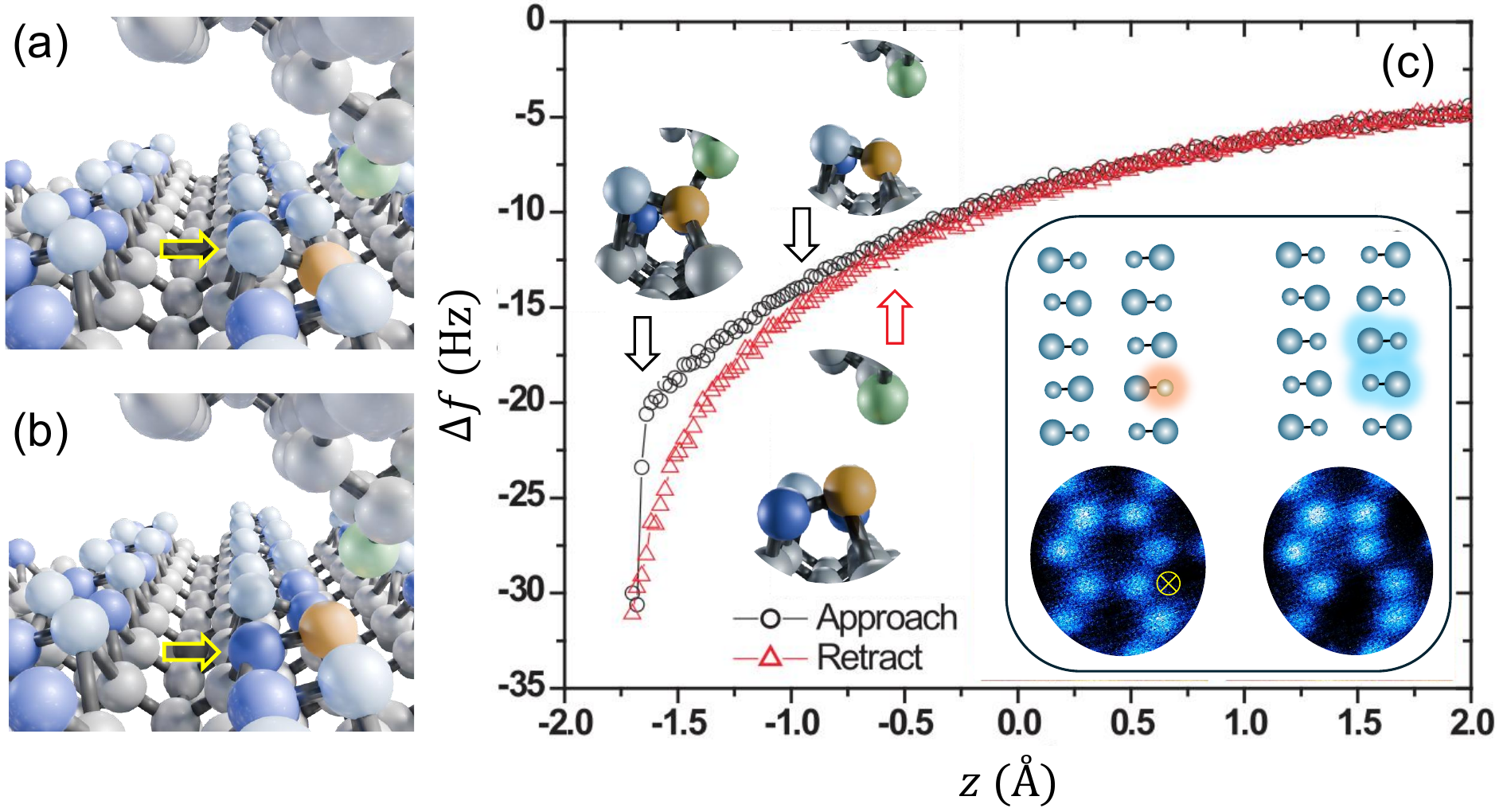}}
\caption{\textbf{Flicking an atomic ``switch''.} A qPlus AFM operating at $\sim$ 5~K (with a nominal sample bias of 0 V) is used to toggle a buckled dimer on the Si(100)-c(4x2) surface. \textbf{(a)},~\textbf{(b)}~are schematic models showing the action of the tip. ``Up'' atoms of dimers are shaded in light blue, ``down'' atoms in dark blue, and the atom targeted by the tip is shaded orange. The atom terminating the tip apex is shaded in green for clarity. The yellow arrows in both (a) and (b) highlight the target dimer before and after the toggle from one state to the other. \textbf{(c)} $\Delta f(z)$ curves acquired during the approach and retraction of the tip. Note, in particular, the sharp discontinuity at a frequency shift of $\sim$ -21 Hz during the approach. This, and the hysteresis observed on retraction, is the signature of a dimer flip event. The inset on the right shows constant frequency shift qPlus AFM images ($\Delta f_{\mathrm{sp}}=-9.1~\text{Hz}$; $V_{\mathrm{bias}}=0$; oscillation amplitude: 250 pm) acquired before and after the flip, with corresponding ball-and-stick schematics of the buckled dimer configurations. (In the ball-and-stick diagrams the targetted atom is also highlighted in orange. Affected dimers after the flip event are highlighted in blue.) The yellow cross-hairs in the qPlus ATM image on the left show the position of the tip at which the $\Delta f(z)$ curves in the main figure were acquired.  In addition to the target dimer, a neighbouring dimer has also flipped (compare the ``before'' and ``after'' images). When starting from the c(4x2) configuration, correlated dimer flipping of this type was always observed, highlighting the strong influence of the local environment.} \label{fig4}
\end{figure*}

Hapala \textit{et al.}'s probe-particle model\cite{Hapala2014,Oinonen2024} reproduces high-resolution AFM contrast -- including apparent intermolecular features of exactly the kind we had observed -- using nothing more than atomic coordinates and empirical pair potentials (not unlike the Girifalco potential discussed in Section 3). Intermolecular electron density does not enter the calculation\footnote{other than indirectly, via equilibrium bond lengths.}. Nonetheless, sharp, ``bond-like'' lines appear due to the response of the flexible probe - be that NTCDI as in our case, CO, or another single-molecule termination -- to the tip-sample energy landscape. Later in 2014, a collaboration between Aalto and Utrecht elegantly demonstrated that intermolecular contrast can indeed arise in the complete absence of any intermolecular bond~\cite{Hamalainen2014}. Their approach was a particularly clever piece of experimental design: rather than relying on a separate control experiment, they chose a molecular system with its control built directly into the structure being imaged. Bis(para-pyridyl)acetylene (BPPA) self-assembles on Au(111) into tetramers held together by genuine \mbox{C-H$\cdots$N} hydrogen bonds between the electronegative pyridinic nitrogens and neighbouring hydrogen atoms. Within each tetramer, however, this same hydrogen-bonding arrangement also forces the nitrogen atoms of two opposing BPPA molecules into close proximity without any hydrogen bond, or bond of any kind, forming directly between them. Strikingly, both sites produced comparable intermolecular contrast features. Returning to C$_{60}$, we also found compelling evidence there~\cite{Jarvis2015} for intermolecular artefacts, in this case over separations considerably larger than in either the NTCDI or BPPA systems, and with a C$_{60}$-terminated tip. This is a van der Waals system with no possibility of hydrogen- or halogen- bonding whatsoever, and is thus about as clean a demonstration as possible that such artefacts are a general feature of dynamic force microscopy\cite{Oinonen2024, Jarvis2015review}\footnote{The term ``non-contact atomic force microscopy'' is often used even when the tip-sample interaction is probed well beyond the equilibrium separation. This usage would appear to have more to do with the technique's historical naming than with the physical regime being accessed in any given measurement. For clarity, and to avoid any ambiguity for readers less familiar with this particular naming convention (or AFM in general), we use the term ``dynamic force microscopy'' here.} and not just a peculiarity of hydrogen-bonded molecules~\cite{Jarvis2015}.

That intermolecular contrast appears in each of these examples despite any direct chemical bond contribution is a sobering result for anyone tempted to read high resolution AFM images too literally. In other words --and to paraphrase the author Douglas Adams\cite{Adams2020BBC}, among others -- it might look like a duck, waddle like a duck, and quack like a duck; yet it can still be a goose. With AFM (and, indeed, scientific imaging in general), the most misleading images are often those that directly confirm our expectations.

\section{Covalent mechanochemistry}
Although the emphasis in previous sections has been on systems underpinned by dispersion forces or hydrogen bonds (be they real or artefactual), stronger interactions -- for one, the repulsive Pauli interaction underpinning high resolution imaging\cite{Gross2009, Moll2012, Jarvis2014Pauli, Brand2019} -- have often been tacitly involved in the measurement process. We now move to a system where those stronger forces are no longer incidental but the entire focus: making, breaking, and manipulating covalent bonds between silicon atoms at the Si(100)-c(4x2) surface. Once again, the AFM tip is used both as actuator and sensor.

Dimers at the Si(100) surface distort via a (pseudo) Jahn-Teller effect (or Peierls distortion)\cite{Verwoerd1980, Robles2012}: electronic degeneracy is removed by a structural distortion (``buckling'') whereby one silicon atom moves up, out of the plane, and the other moves down. This leads to a bond angle of approximately 19$^{\circ}$ across the dimer and a concomitant charge transfer from the ``down'' to the ``up'' atom. As such, and assuming that an appropriate mechanochemical actuation protocol can be found, a dimer represents the smallest conceivable in-plane toggle switch. In Fig.~4 we test this hypothesis directly: can a single dimer be mechanochemically toggled between its two buckled configurations, on demand, using nothing but the force exerted by an AFM tip\cite{Sweetman2011}? As Fig. 4(c) illustrates, it is indeed possible to flip dimer orientation (and, although not shown in the figure, to also restore the original buckled state\cite{Sweetman2011}). 

While the AFM-actuated ``toggle switch'' protocol therefore works in practice, there is an important proviso: across hundreds of manipulation events, we found that it was not possible to flip a single dimer in the native c(4x2) configuration without influencing a neighbouring dimer\cite{Sweetman2011,Sweetman2011PRB}. In other words, only correlated flips were possible. This, and the variation in the propensity for a dimer to flip depending on its proximity to defects \cite{Sweetman2011PRB}, underscore that both the local and non-local environment underpin the energy landscape and dynamics\cite{Sweetman2011,Sweetman2011PRB}. \textit{Ab initio} simulations using DFT also highlight that orbital orientation and charge density at the apex of the tip, in concert with the precision of probe positioning, determine the probability of not only dimer flips but a variety of other types of atomic manipulation \cite{JarvisPRB2012, Jarvis2013Beilstein}. The choice of terminating geometry (e.g. Si(111)-like, Si(100)-like...), for one, can preclude certain classes of manipulation event while enabling others. We also note that Fig. 4 focuses on dimer switching via interactions in the attractive regime of the tip-sample potential: toggling a dimer via ``pushing'' rather than ``pulling'' was not possible for the clean Si(100) surface. Later work by Sweetman \textit{et al.}\cite{Sweetman2020} on Pb dimers on Si(100), however, demonstrated that repulsive tip-sample interactions in that system \textit{could} actuate flip events. 

\section{Bottom up, from the top} Although the dimer flipping of the previous section is a form of what is known in the SPM community as vertical manipulation, both the silicon atom at the tip apex and those comprising the dimer remain firmly back-bonded to their neighbours throughout (not least due to the strength and directionality of the covalent bonds). This is a very different mechanism as compared to, for example, the fascinating atomic interchange process reported by Sugimoto \textit{et al.}\cite{Sugimoto2008} for the Sn/Si-($\sqrt{3}\times\sqrt{3})\text{R}30^{\circ}$ surface, whereby an atom at the surface is exchanged for an atom at the tip, repeatedly, enabling atomically precise patterns of one element (in this case, silicon) to be written into a matrix of the other (tin). 

While the level of control here is undoubtedly impressive, the Sn/Si system is somewhat ``bespoke''; there is a fortuitous combination of potential energy landscape and chemistry that is not readily transferable to other materials. Decades before Sugimoto \textit{et al}'s observation and exploitation of the interchange mechanism, atom transfer between tip and surface involving what might be described as more ``direct'' routes had been demonstrated by a number of groups, including the landmark Xe single atom switch paper by Eigler, Lutz, and Rudge\cite{EiglerSwitch}. The authors highlight both Becker \textit{et al.}'s\cite{Becker1987} and Lyo and Avouris'\cite{Lyo1991} ``prior art'' on tip$\leftrightarrow$sample atom transfer for Ge and Si surfaces, respectively. Decades after both Lyo and Avouris' formative paper on vertical manipulation at the Si(111) surface and Shen~\textit{et al.}'s field-defining work on H atom removal from hydrogen-passivated Si(100)\cite{Shen1995}, and within months of publication, both Pavli{\v c}ek \textit{et al.}\cite{Pavlicek2017-hh} and Huff \textit{et al.}\cite{Huff2017} each achieved mechanochemical transfer of a hydrogen atom from a qPlus AFM tip to a silicon dangling bond on H:Si(100)-(2x1). Huff \textit{et al.} coined the term ``atomic whiteout'' for this process as it represents an essential contribution to the atomic manipulation toolbox: error correction. (During our own qPlus AFM studies of H:Si(100)-(2x1)\cite{Sharp2012}, we had attempted to achieve exactly this type of H transfer to silicon dangling bonds via qPlus AFM, with no success. We were therefore particularly well placed to appreciate the significant difficulty of what Pavli{\v c}ek \textit{et al.}\cite{Pavlicek2017-hh} and Huff \textit{et al.}\cite{Huff2017} accomplished.)  

Throughout the scanning probe literature, however, atomic or molecular manipulation via vertical exchange or transfer is much less common than lateral translation of adsorbates when it comes to assembly of atomically precise nanostructures. Since its inception by Eigler and Schweizer\cite{Eigler1990}, lateral manipulation by pushing, pulling, and/or sliding adsorbates has led to an astoundingly wide variety of custom-built, atomically precise arrangements of matter that have often underpinned key advances in control at the quantum level\cite{Crommie1993, Manoharan2000, Khajetoorians2019, Sierda2023}. It is easy to understand why lateral, rather than vertical, manipulation is much more prevalent. It boils down to the usual suspect: the tip. 

Transfer of an atom to or from the tip can often be a significantly more troublesome and fraught process for a probe microscopist than lateral manipulation. A well-worn protocol to coerce tips into yielding atomic, molecular, or submolecular resolution -- or to remove double-, triple-, or multiple-tip artefacts -- is to gently (or not so gently) indent the apex into the underlying surface. In other words, tip-sample contact is very deliberately exploited to change the atomistic or mesoscopic structure of the apex of the probe. Retaining the tip state can therefore be a considerable challenge in vertical manipulation events where the aim is to deposit or extract atoms. This is one of the reasons why Sugimoto \textit{et al.}'s vertical interchange protocol\cite{Sugimoto2008} is particularly impressive: the tip state, both with regard to maintaining atomic resolution and its chemical/elemental nature, is continually restored. This type of ``recyclable'' tip is particularly sought after in the SPM community\cite{AbbasiPerez2021, Custance2009}, and we return to this point below.

\begin{figure}[b!]
\centerline{\includegraphics[width=0.5\textwidth]{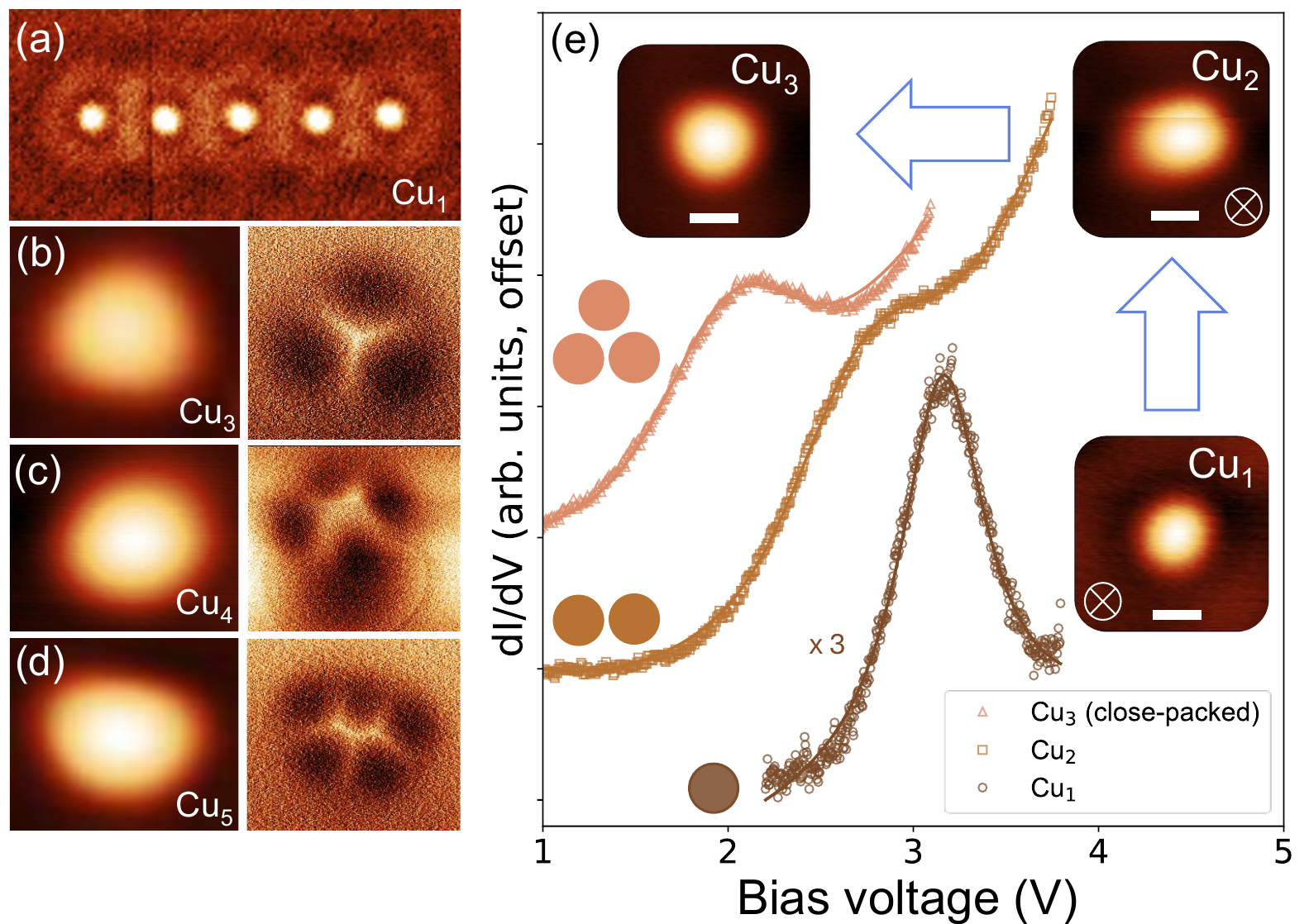}}
\caption{\textbf{Adding a dimension to atom-by-atom assembly.} Deposition of copper atoms and transfer/formation of Cu$_n$ clusters using a qPlus sensor in combined STM/AFM mode. \textbf{(a)}~Constant-current STM image (-20 mV, 100 pA) acquired following sequential deposition of single atoms from the tip. Prior to the deposition, the tip was ``charged'' by indenting it 5 nm into the Cu(111) surface with an applied bias of -1 V. Each atom was subsequently deposited via 6.8\AA~indents into the surface from the tip height defined by the setpoint condition. \textbf{(b),~(c),~(d)}~STM~images and subsequent constant-height frequency-shift images using a CO-functionalised apex, (left and right, respectively) of \textbf{(b)}~a~Cu$_3$ cluster deposited from the tip using a similar protocol to that for (a); \textbf{(c)}~a~Cu$_4$ cluster synthesized by joining a Cu atom to a linear Cu$_3$ trimer using a lateral manipulation strategy similar to that described by F\"olsch and co-workers\cite{Folsch1, Folsch2, Folsch3}; \textbf{(d)}~a~Cu$_5$ cluster deposited directly from the tip. \textbf{(e)~Atom-by-atom synthesis of a close-packed Cu$_3$ trimer using vertical manipulation.} Resonances in dI/dV spectra have peak energies dependent on the value of $n$ and thus can be used to determine cluster size in the absence of a CO-functionalised tip\cite{Folsch1, Folsch2, Folsch3}. The close-packed trimer was fabricated by first forming a dimer and then a trimer by vertical manipulation, depositing an atom from the tip in each case (blue arrows denote the sequence). The cross-hairs in the insets indicate the lateral tip position during the deposition process. (Sample bias:-20 mV; $\langle I_{\mathrm{t}}\rangle=1$~pA; oscillation~amplitude:~400$\pm 20$ pm at the setpoint conditions. We highlight, however, that during each Cu transfer event the damping channel saturated and the oscillation amplitude was no longer held constant). Scale bars: 0.5 nm.}
\end{figure}

A more broadly applicable route to controlled vertical manipulation comes from the point-contact literature, where there is a substantial body of work on simultaneous measurement of force and conductance during the formation and rupture of atomic-scale metal contacts. This spans, to list just a few exemplars, gold nanocontacts at room temperature~\cite{Rubio1996}, platinum and copper at $\sim 5~\text{K}$ ~\cite{Ternes2011}, point contacts to C$_{60}$~\cite{Hauptmann2012, Brand2019}, and, strikingly, the transfer of a single Ag atom from a metal-terminated tip directly into~\cite{Sperl2011}, and out of\cite{Sperl2011JACS}, the macrocycle of an adsorbed phthalocyanine molecule -- a single-atom analogue of on-surface metalation. 

Our interest in atom transfer stems from a long-standing aspiration to add a dimension to atom-by-atom assembly: to combine lateral manipulation with vertical transfer so as to ultimately form three-dimensional nanoclusters. The previously unpublished data of Fig. 5 shows first steps in this direction, directly inspired by the elegant Cu-on-Cu(111) atomic manipulation work of Stefan F\"olsch and co-workers at the Paul-Drude-Institut f\"ur Festk\"orperelektronik in Berlin\cite{Folsch1, Folsch2, Folsch3}. We first ``load'' the tip by embedding it in the Cu(111) surface by a few nanometres at a bias of order 1~V. Atoms and small Cu$_n$ clusters can then be deposited from the tip by indentation\cite{SawReview}. In the case shown in Fig. 5(a), where five Cu atoms have been sequentially deposited in STM mode, we pushed the tip towards the surface by 6.8~\AA~ in each case (from a $z$ height set by the setpoint tunnelling conditions, namely -20 mV and 100 pA). We should stress that, in our experience, reliable, consecutive single-atom deposition of the type shown in Fig. 5(a) tends to be the exception rather than the norm. Generally, single atom deposition will often be followed by transfer of a Cu$_n$ cluster, be that a dimer, trimer or something larger. (Other groups\cite{NanosurfLab} have demonstrated much more consistent single atom transfer over hundreds of events.) Figs. 5(b) and (d) show STM and qPlus AFM images acquired using a CO-functionalised tip\cite{Bartels1997,Gross2009} -- the latter using the strategy of Emmrich \textit{et al.}\cite{Emmrich2015} -- of Cu$_3$ and Cu$_5$ deposited directly from the tip; Fig. 5(c), on the other hand, is a Cu$_4$ cluster formed by joining a Cu atom to a linear Cu$_3$ trimer using lateral manipulation\cite{Folsch1, Folsch2, Folsch3}. 

It is the data shown in Fig. 5(e), however, that is of particular relevance to the vertical assembly goal. Although CO-termination -- or, as we have seen in previous sections, other single-molecule functionalisation -- is a powerful method of providing direct, real-space sub-cluster contrast to the level whereby the constituent atoms can be counted, switching back and forth between a Cu-terminated and CO-functionalised apex is an exceptionally tedious and time-consuming process. During deposition and assembly of Cu$_n$ clusters we therefore rely on the $n-$dependent dI/dV resonances identified and exploited by F\"olsch and co-workers\cite{Folsch1, Folsch2} for identification, only switching to a CO tip following assembly. Note in Fig. 5(e) how the peak in the dI/dV spectrum progressively shifts to lower energy\cite{Folsch1} as a function of $n$. 

Starting with Cu$_1$, and operating in dynamic STM mode with an oscillation amplitude of 400 pm ($\pm$20 pm at the height set by the setpoint conditions of -20 mV and $\langle I_{\mathrm{t}}\rangle=1~\text{pA}$), we positioned the tip at the position marked by the cross-hairs in the inset on the lower right of Fig. 5(e). We then disengaged feedback and acquired $\Delta f(z)$, $I(z)$, $A_{\mathrm{osc}}$, and damping($z$) curves for a tip excursion of +2nm to -0.54 nm, where the positive direction is away from the surface and the zero of the excursion is set by the feedback stabilisation condition. Note that the tip position shown in the inset image is to one side of the atom. Repeated attempts where the tip was positioned directly on top of the underlying Cu atom caused only lateral displacement; the off-centre position of the tip was critical for successful atom deposition. When we re-scanned the area following the tip excursion, the Cu$_1$ had ``converted'' to a Cu$_2$ dimer; the discontinuity in the image is characteristic of dimer diffusion\cite{Folsch1, Repp2003} and there is a significant change in the lineshape and position of the peak in the dI/dV spectrum. (We note in passing that although $\Delta f(z)$ and $I(z)$ curves were acquired, their interpretation is challenging due to the large excursions of the amplitude and damping at closest approach during atom transfer.)

\begin{figure*}[b!]
\centerline{\includegraphics[width=1.0\textwidth]{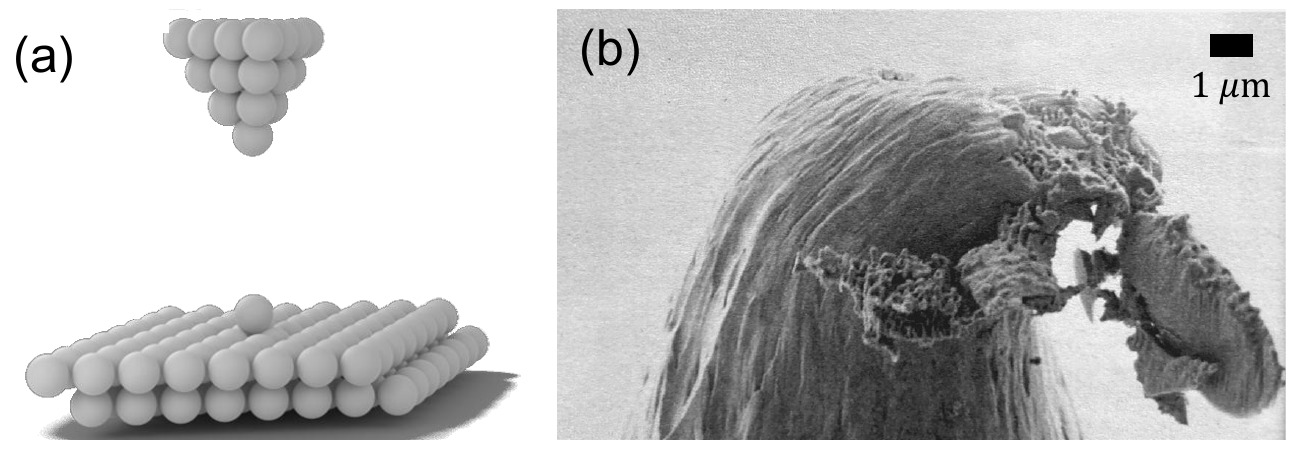}}
\caption{\textbf{Model tips.} On the left, an artist's impression, generated using the Blender application, of an idealised SPM tip above a metal surface. On the right, a scanning electron microscope (SEM) image of a tip that was used for prolonged periods of both atomic resolution imaging and high precision manipulation of Cs atoms on InSb(110). From the perspective of the probe microscopist, if not the electron microscopist, what is shown in (b) was therefore in many ways a model tip apex. Shortly after the acquisition of the SEM image, the tip was machined to a more presentable state via focussed ion beam milling. This restored a better-defined microscopic form (albeit with no guarantee of a similarly well-behaved mini-tip right at the apex.) }
\end{figure*}

Having formed the Cu$_2$ dimer, we positioned the tip again off centre (inset in upper right of Fig. 5(e)) and again measured $\Delta f(z)$ and $I(z)$ curves, with a slightly smaller range: +2~nm to -0.48~nm. (For each deposition, we progressively increased the piezo extension until discontinuities and hysteresis were observed in the frequency shift and tunnel current curves.) This resulted in the formation of a close-packed Cu$_3$ cluster, as shown in the dI/dV spectrum and image in the upper left corner of Fig. 5(e). What is particularly interesting about this result, beyond the assembly of a cluster via vertical atom transfer, is that it is not possible to build a close-packed trimer in this way using atom-by-atom lateral manipulation. We found, entirely in line with F\"olsch \textit{et al.}'s observations\cite{Folsch1,Folsch2}, that adding a Cu atom to a Cu dimer via lateral manipulation invariably resulted in a linear trimer. Vertical manipulation has therefore enabled a pathway across the tip-sample energy landscape that is essentially prohibited using lateral translation alone. We have some tentative evidence that vertical addition of another Cu atom to the close-packed Cu$_3$ trimer can result in a non-planar Cu$_4$ cluster -- as distinct from the planar Cu$_4$ shown in Fig. 5(c) which was synthesized by adding a Cu atom to a linear Cu trimer via lateral manipulation -- but this is very much work in progress.

\section{Tipping the balance}
In textbooks and scientific papers, SPM tips are generally schematically illustrated as perfect single-atom or single-molecule terminations; the ray-traced representations in Fig. 2, Fig. 4, and Fig. 6(a) of this paper are good examples. Probe microscopists, however, know full well that these idealisations are exactly that: ideal, model representations of the tip. BQG graciously acknowledged the inspiration and influence of John Pethica (then at Cambridge, now at Trinity College Dublin and Oxford) on the development of the AFM, via his realisation of the substantial tip-sample forces at play in the STM. (In a fascinating first-hand account of the birth of AFM (and STM), Christoph Gerber also notes Pethica's influence\cite{GerberKavli}.) Years later, Pethica memorably highlighted -- during the UK SPM meeting in 2009 at the National Physics Lab, if memory serves -- a related and particularly thorny problem when it comes to SPM: tip artefacts and convolution. He pointed out that a guilty secret of all probe microscopists relates to the stacks of unexplained images languishing at the bottom of filing cabinet drawers or on neglected sectors of hard drives. In each case, the interpretation of the image was stymied by lack of knowledge of the precise tip structure.

Although inverse imaging\cite{Giessibl2000,Schull2009,Schull2011, Lakin2013} (as demonstrated in Fig. 2(a)) can be used to characterise the tip, this is a relatively niche protocol that is possible only when the effective tip radius of curvature exceeds that of the proxy ``probe'' at the surface. Moreover, while this approach can directly image the tip apex, it provides little, if any, information on the broader nanoscopic, mesoscopic, or, indeed, microscopic structure of the tip. As Fig. 6 shows, those can be somewhat less ideal than one might hope. Despite the rather mangled appearance of the tip in Fig. 6(b), it not only yielded high quality atomic resolution images of the InSb(110) lattice and of Cs atoms adsorbed on that surface, but was used for atomically precise manipulation of Cs. This, of course, relates to the local nature of the interactions; for STM, the exponential dependence of tunnelling on tip-sample separation means that all it takes is for one atom to be sticking out fractionally more than the rest and this will dominate current flow. However, not only can there be more than one ``mini-tip'' of this type, but the tunnelling centre can shift during scanning. Add in the detection of force gradients via, for example, a qPlus sensor, and a plethora of different hybrid tip artefacts and convolutions can result.

The consequences of this inhomogeneity for a given measurement are often far from straightforward to isolate (hence Pethica's comment about perplexing images being squirreled away in filing cabinets). In very recent work currently under review\cite{Moller2026}, we have interpreted the intermittent, "bursty" tunnelling current signal produced by perylenetetracarboxylic dianhydride (PTCDA) molecules diffusing under the tip as being underpinned by the heterogeneity of the probe-mediated potential energy landscape. The distribution of times between successive tunnelling pulses follows the same truncated power-law form as has been reported for a variety of activities on substantially larger length- and time-scales\cite{Barabasi2005} and which have been interpreted in terms of ``universal'' scale-free dynamics. Kinetic Monte Carlo simulations, in concert with our measurements of the tunnel current fluctuations, however, indicate that the characteristic heavy-tailed statistics in the molecular diffusion case instead emerge naturally from the spatially heterogeneous energy landscape imposed by the tip\cite{Moller2026}. 

Although the combined analysis of tunnelling current fluctuations and KMC simulations can provide key insights into the energy landscape, this is an indirect route to ascertaining the influence of tip heterogeneity. Without accompanying AFM measurements (which were not possible in the PDCTA/Ag(110) case\cite{Moller2026}), it is challenging, at best, to directly probe the influence of the inhomogeneity at the tip apex on the forces experienced by the diffusing molecules. With a combined AFM-STM system, however, the frequency shift of the qPlus sensor or cantilever provides a natural measure of the magnitude of the integrated vdW background due to the finite radius of curvature of the tip. In other, as yet unpublished, work\cite{Carlisle2026NCAFM}, we found that even at temperatures as low as 330~mK it was impossible to non-perturbatively image a dibromoterfluorene (DBTF) molecule on Ag(111)\cite{Donato1} without first reducing the background dispersion forces to a minimal level (via repeated tip sharpening through controlled indentation, monitored by the reduction in the magnitude of the total frequency shift.)       

\section{AFM ex machina}
Be it dispersion forces, covalent bonds, or Pauli repulsion-mediated interactions that underpin the imaging mechanism, we've seen throughout this retrospective that AFM is shaped just as much by the tip as the sample (unsurprisingly for a force microscope, given Newton's third law\cite{Bader2000}.) In this closing section, we look to the coming decade (and beyond) and ask not only to what extent artificial intelligence can assist with post-acquisition data classification and analysis, but how it might be integrated \textit{in situ} to circumvent, or at least tame, the vagaries of the tip. With appropriate AI-mediated apex control -- including, critically, error correction protocols -- in place, could atom-by-atom assembly on mesoscopic, microscopic, or even macroscopic scales be achieved? Could an AI be trained to select an appropriate AFM tip state for specific types of atomically precise mechanochemical manipulation? In its most ambitious framing, this is a question about how close we can come to the kind of molecular nanofactory technology posited, in various guises, by Drexler more than three decades ago\cite{Drexler1992}.

Realising anything even remotely resembling atomic precision manufacturing, however, demands much more than manipulation protocols alone: it requires an entire supporting architecture of autonomous decision-making, from recognising what is being imaged to deciding what to do next. Kalinin and co-workers have been especially prolific and influential at the interface of nanoscience and AI in this regard, developing autonomous and active-learning-driven workflows for both scanning probe and electron microscopy that make real-time, in-the-loop decisions about where and how to acquire data~\cite{Kalinin2015, Rickman2019, Kalinin2022, Madika2025}. In work in a somewhat similar vein, our group at UoN, in collaboration with Ingmar Swart's team at Utrecht, trained convolutional neural networks (CNNs) to classify STM images of metal and semiconductor surfaces\cite{Gordon2019}. Our ML-related work subsequently focused on characterising the state of the tip apex in real time from small numbers of line-scans (as is common for human probe microscopists) ~\cite{Gordon2020a}, searching large repositories of AFM images for self-organised nanostructures of interest~\cite{Gordon2020b}, and developing CNN-based approaches to image segmentation\cite{Farley2020}.  

Valuable as these classification and analysis strategies can be, they address only one side of the autonomous-microscopy problem, \textit{viz.} what are we looking at? The other side, rather more germane to the "assembler" framing above, relates to just how we controllably manipulate what's in the proximity of the tip. Automated, atom-by-atom assembly of pre-defined nanostructures is not itself a new idea. Celotta, Stroscio and co-workers\cite{Celotta2014} demonstrated impressive autonomous STM-driven assembly of atomically perfect nanostructures over a decade ago, followed soon after by Kalff \textit{et al.}'s atomic memory prototype involving many thousands of automated atomically-precise manipulation steps\cite{Kalff2016}, both well before deep-learning-driven approaches were in vogue. What has changed in the intervening years is the sophistication of the control policy governing each manipulation step. Reinforcement learning, in particular, has been used, remarkably, to learn the precise tip trajectory needed to extract molecules from a molecular lattice one at a time \cite{Leinen2020}, to control the position and orientation of a single dipolar molecule via the electric field of an STM tip~\cite{Ramsauer2023}, and to manipulate individual Ag adatoms on Ag(111) with atomic precision~\cite{Chen2022}. In each case, the manipulation policy was learned directly from continuous experimental feedback rather than relying on a fixed, hand-tuned protocol.

Tip control specifically, as opposed to the manipulation or assembly process more broadly, has also received some attention from the AI/ML perspective. We explored this ourselves some time ago -- in the AI dark ages of 2011 -- using a genetic algorithm to ``evolve'' the imaging capability of the tip, without human intervention~\cite{Woolley2011}. More recently, Alldritt and co-workers (Aalto) have developed a machine-learning-classifier-driven pipeline for automated CO tip functionalisation~\cite{Alldritt2022}, while Rashidi and Wolkow have demonstrated autonomous, ML-guided \textit{in situ} tip conditioning in the context of dangling bond architecture, including logic gate generation, on hydrogen-passivated Si(100) surfaces~\cite{Rashidi2018}, while Wang \textit{et al.}\cite{Wang2021} have adopted a similar ML-directed tip preparation approach for tunnelling spectroscopy.

To date, machine-learning-facilitated atomic manipulation has concentrated on scanning tunnelling microscopy-based protocols, not least because STM has been the tool of choice for the majority of single atom/molecule positioning demonstrations. However, and as is hopefully abundantly clear from the preceding sections, atoms and molecules ultimately move in response to local forces. With this in mind, we have focussed most recently on qPlus AFM-driven manipulation of Cs adatoms across an InSb(110) surface\cite{Sierda2023}. Following Ternes \textit{et al.}\cite{Ternes2008}, we capture both vertical and lateral forces at the moment an atom moves (Fig. ~7). As cesium adatoms on InSb(110) are electrically charged and the substrate is weakly conducting, electrostatic interactions play a particularly important role in the manipulation dynamics. Currently we are combining large, multivariate datasets -- including probe–sample forces, energy dissipation, and tunnelling current -- with reinforcement learning\cite{Leinen2020, Chen2022} with a focus on discovering effective strategies for atomic manipulation. By providing the RL agent with a richer view of each manipulation event via a set of complementary experimental observables, our aim is to not only improve the efficiency of the (automated) atomic manipulation process but to gain insights into the physics and physical chemistry underpinning the tip-mediated translation of adsorbates.

\begin{figure}[t!]
\centerline{\includegraphics[width=0.5\textwidth]{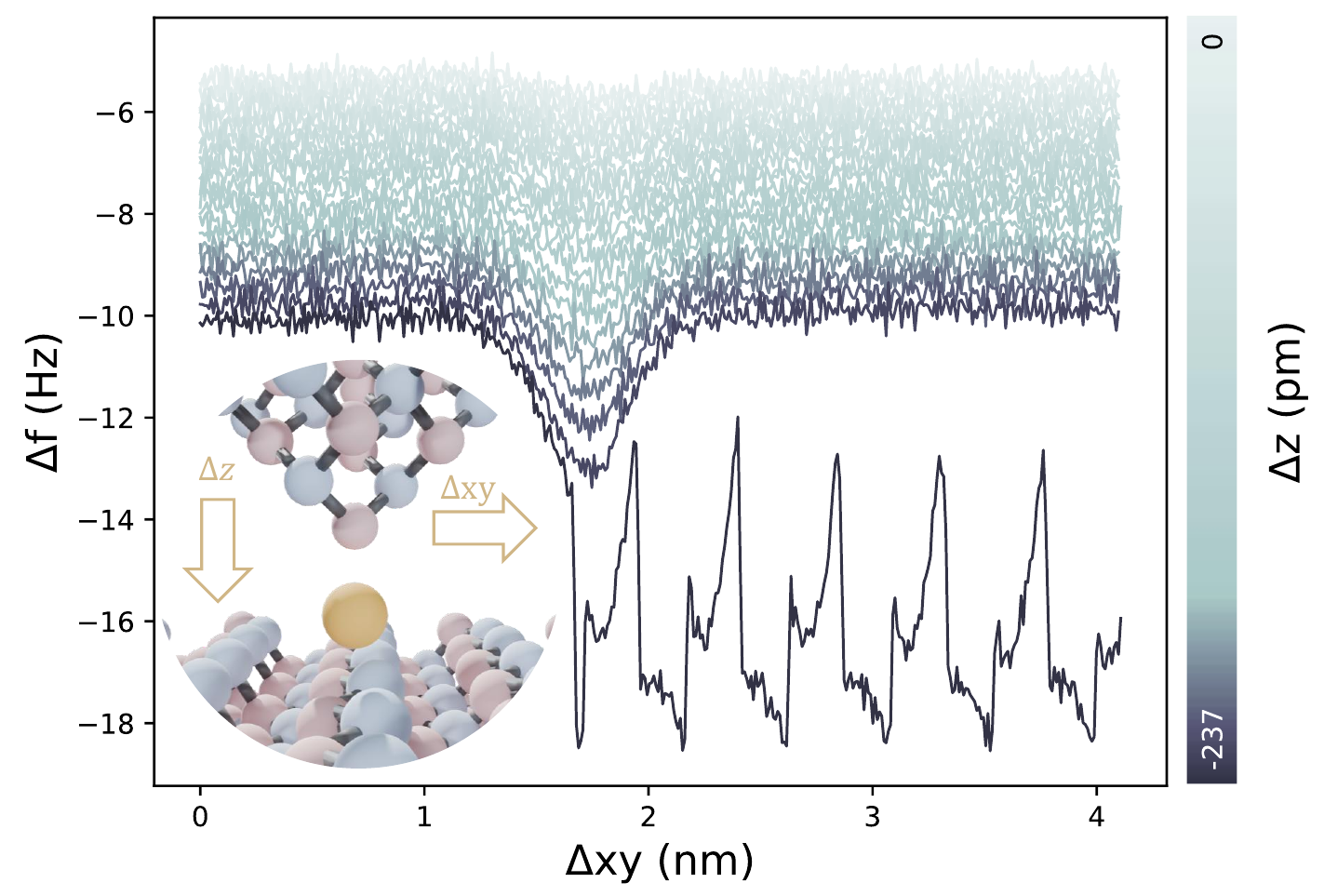}}
\caption{\textbf{Site by site. AFM-directed manipulation of a Cs atom on InSb(110).} A stack of profiles of frequency shift ($\Delta f$) vs lateral tip position ($\Delta (xy)$) as a function of the displacement of the qPlus AFM tip towards the surface ($\Delta z$). At a threshold tip height, the Cs atom moves in discrete steps across the underlying InSb(110) lattice. \textbf{Inset:} Schematic showing the Cs atom (gold) and the InSb(110) substrate (blue/red).} 
\end{figure}

\section{Conclusions}
Looking back forty years on, BQG's appraisal of the capabilities of their then-newly-invented instrument\cite{Binnig1986} was both prescient and, in hindsight, conservative. (As Christoph Gerber has put it\cite{GerberKavli}, given that an AFM has already been sent to the Martian surface -- difficult to predict back in 1986 -- one could argue that not even the sky is the limit!) Throughout this review we have focussed not only on the span of measurable interaction strengths first predicted by Binnig, Quate, and Gerber but on what might be best termed the ``controlled invasiveness'' of the component at the core of their microscope: the tip. Although every measurement is, at some level, a convolution of object and instrument properties, AFM -- and SPM in general -- take this to an extreme; when operating at, or beyond, the equilibrium tip-sample separation it is clear that the probe can no longer be a non-invasive, passive observer. While this is obvious for the contact- or near-contact measurement regime, we have also aimed to highlight that, despite its operation in the near field, the probe's influence on a measurement is not always confined to the well-defined handful of atoms at its very apex. Real tips are heterogeneous on the nanoscopic, mesoscopic, and microscopic scales, and that heterogeneity leaves a measurable imprint even on ostensibly ``gentle'', non-invasive measurements\cite{Moller2026}. Robustness to exactly this kind of imperfection will be a particularly challenging target for future AI-assisted probe microscopy.

There is a pleasing resonance here, however, with the thinking of one of the AFM's own inventors. In \textit{Aus dem Nichts}\cite{BinnigBook}, Gerd Binnig argues that genuine creativity arises not from tidy, structured processes but from disorder itself, mirroring the mutation and selection pressures found throughout biological evolution. Having spent this review documenting results that often hinge on the probe's own inhomogeneity and multi-valley potential energy (or ``fitness'') landscape, it is difficult not to read Binnig's thesis as bearing directly on the AFM discipline he, with Quate and Gerber, founded. 

%\backmatter
\bmsection*{Author contributions} This is a review, retrospective, and perspective article. As such, the author list largely comprises those who were responsible for the original work cited throughout the paper. We acknowledge a productive, stimulating, and lengthy collaboration with Uwe Thiele (now at the University of M\"unster) in relation to the results discussed in Section 2 of the paper on far-from-equilibrium pattern formation in nanoparticle assemblies. 

For the previously unpublished data of Fig.~5 and Fig.~7, the author contributions are as follows. \textbf{Fig.~5:}~FJ,~CF,~and~PM carried out the experiments and were responsible for data analysis; PM drafted the description of the results; ~\textbf{Fig.~7:}~The overall reinforcement learning framework, including coding, is due to SAP; experiments: SAP, OP; analysis and strategy: SAP, OP, PM, and BK. BK drafted the description of the results. The paper as a whole was drafted with input and feedback from co-authors. 

\bmsection*{Acknowledgments}
We thank the editors for the invitation to contribute to the \textit{Forty Years of Atomic Force Microscopy} special issue. We also thank Mike Fahy and Chris Parmenter of UoN's nanoscale and microscale Research Centre (nmRC) for electron microscopy and focussed ion beam milling in relation to Fig. 6. None of the work described here would have been possible without the expert technical support (mechanical, electronic, ultrahigh vacuum, cryogenic) in our workshops; we thank Pete Smith, Nick Botterill, Andrew Solomon, Ian Taylor, Matt Woolley, and Sanjeev Taak in particular. We also gratefully acknowledge financial support from a variety of publicly-funded organisations, charities, and professional bodies including the Engineering and Physical Sciences Research Council (EPSRC), Leverhulme Trust, Royal Society, and EU Framework Programmes 5, 6 and 7. The corresponding author (PM) is particularly grateful to EPSRC (and, by extension, UK taxpayers) for the award of a Leadership Fellowship (2008 -- 2014; EP/G007837/1) and an Established Career Fellowship (2020 -- 2026, EP/T033568/1). Both fellowships underpinned much of the work described in this review. 3D models and visual renders included in the figures were generated using Blender v5.2 (Blender Foundation, Amsterdam, Netherlands), an open-source 3D creation suite available at \href{https://www.blender.org}{https://www.blender.org}.\\
\indent There are a number of researchers who contributed to the work reviewed in this paper but for whom we could not locate current valid email addresses and, as such, they unfortunately could not be listed as co-authors. We would like to warmly acknowledge the contributions of those researchers: Joesph Bamidele (former PhD student at Kings College London), Cristina Chiutu (former PhD student, University of Nottingham), Rosanna Danza (former PhD student, University of Nottingham), Andrew Lakin (former PhD student, University of Nottingham), and Michael Taylor (former PhD student, University of Nottingham).\\

\noindent \textbf{List of present addresses}\\
Matthew O Blunt: Department of Chemistry, University College London (UCL), London WC1H 0AJ\\
Neil R. Champness: School of Chemistry, University of Birmingham, Edgbaston, Birmingham, B15 2TT, UK \\
Subhashis Gangopadhyay: Department of Physics, Birla Institute of Technology and Science (BITS) Pilani, Pilani 333031, Rajasthan, India \\
James Hayton: PhD academy Ltd.\\
Samuel P Jarvis: School of Physics and Astronomy, Lancaster University, Lancaster LA1 4YB\\
Natalio Krasnogor: School of Computing, 1 Science Square, Newcastle upon Tyne NE4 5TG, UK \\
Adrian Radocea: Varda, 225 S Aviation Blvd, CA 90245, USA\\
Philipp Rahe: Molecular Quantum Structures, Fachbereich Mathematik/Informatik/Physik, Universit\"at Osnabr\"uck, Raum 32/215a, Barbarastrasse 7, 49076 Osnabr\"uck, Germany\\
Mohammad Abdur Rashid: Department of Physics, Jashore University of Science and Technology, Jashore 7408, Bangladesh\\
Hongqian Sang: State Key Laboratory of Precision Blasting, Jianghan University, Wuhan, 430056, China\\
Andrew Stannard: Department of Chemistry, Imperial College London, London W12 0BZ, United Kingdom \\
Julian Stirling: Humanitarian Technology Trust, Bath, UK\\
Adam Sweetman: School of Physics \& Astronomy, University of Leeds, Leeds, LS2 9JT, UK\\

\bmsection*{Conflict of interest}
The authors declare no potential conflict of interest, financial or otherwise.

\bmsection*{Data Availability} Although the figures in this paper have not been previously published in the form shown here, the majority of the data discussed herein is associated with earlier publications. The exceptions are Fig.~5 and Fig.~7. Raw data associated with those figures is available at \href{https://doi.org/10.17639/nott.40012}{https://doi.org/10.17639/nott.40012}.

\bibliography{FortyYearsAFM}

\begin{thebibliography}{100}

\bibitem{Binnig1986}
Binnig G, Quate CF, Gerber C.
\newblock Atomic Force Microscope.
\newblock Phys Rev Lett. 1986 Mar;56:930--933.
\newblock Available from:
  \url{https://link.aps.org/doi/10.1103/PhysRevLett.56.930}.

\bibitem{Martin1987}
Martin Y, Williams CC, Wickramasinghe HK.
\newblock Atomic force microscope–force mapping and profiling on a sub
  100‐Å scale.
\newblock Journal of Applied Physics. 1987 05;61(10):4723--4729.
\newblock Available from: \url{https://doi.org/10.1063/1.338807}.

\bibitem{Moriarty2002}
Moriarty P, Taylor MDR, Brust M.
\newblock Nanostructured Cellular Networks.
\newblock Phys Rev Lett. 2002 Nov;89:248303.
\newblock Available from:
  \url{https://link.aps.org/doi/10.1103/PhysRevLett.89.248303}.

\bibitem{OShea2002}
O'Shea JN, Phillips MA, Taylor MDR, Moriarty P, Brust M, Dhanak VR.
\newblock Colloidal particle foams: Templates for Au nanowire networks?
\newblock Applied Physics Letters. 2002 12;81(26):5039--5041.
\newblock Available from: \url{https://doi.org/10.1063/1.1526924}.

\bibitem{Martin2004}
Martin CP, Blunt MO, Moriarty P.
\newblock Nanoparticle Networks on Silicon:  Self-Organized or Disorganized?
\newblock Nano Letters. 2004 10;4(12):2389--2392.
\newblock Available from: \url{https://doi.org/10.1021/nl048536w}.

\bibitem{Hayton2007}
Hayton JA, Pauliac-Vaujour E, Moriarty PJ.
\newblock Anisotropic assembly of colloidal nanoparticles: Exploiting substrate
  crystallinity.
\newblock Nano. 2007;02(06):361--365.
\newblock Available from: \url{https://doi.org/10.1142/S1793292007000714}.

\bibitem{Blunt2007}
Blunt MO, Martin CP, Ahola-Tuomi M, Pauliac-Vaujour E, Sharp P, Nativo P,
  et~al.
\newblock Coerced mechanical coarsening of nanoparticle assemblies.
\newblock Nat Nanotechnol. 2007 Mar;2(3):167--170.

\bibitem{Brust1994}
Brust M, Walker M, Bethell D, Schiffrin DJ, Whyman R.
\newblock Synthesis of thiol-derivatised gold nanoparticles in a two-phase
  Liquid–Liquid system.
\newblock Journal of the Chemical Society, Chemical Communications. 1994
  01;30(7):801--802.
\newblock Available from: \url{https://doi.org/10.1039/C39940000801}.

\bibitem{Ge2000}
Ge G, Brus L.
\newblock Evidence for Spinodal Phase Separation in Two-Dimensional Nanocrystal
  Self-Assembly.
\newblock The Journal of Physical Chemistry B. 2000 09;104(41):9573--9575.
\newblock Available from: \url{https://doi.org/10.1021/jp002280a}.

\bibitem{Rabani2003}
Rabani E, Reichman DR, Geissler PL, Brus LE.
\newblock Drying-mediated self-assembly of nanoparticles.
\newblock Nature. 2003 Nov;426(6964):271--274.

\bibitem{Sztrum2006}
Sztrum Câ, Rabani E.
\newblock Out-of-Equilibrium Self-Assembly of Binary Mixtures of Nanoparticles.
\newblock Advanced Materials. 2006;18(5):565--571.
\newblock Available from:
  \url{https://advanced.onlinelibrary.wiley.com/doi/abs/10.1002/adma.200501408}.

\bibitem{Martin2007}
Martin CP, Blunt MO, Pauliac-Vaujour E, Stannard A, Moriarty P, Vancea I,
  et~al.
\newblock Controlling Pattern Formation in Nanoparticle Assemblies via Directed
  Solvent Dewetting.
\newblock Phys Rev Lett. 2007 Sep;99:116103.
\newblock Available from:
  \url{https://link.aps.org/doi/10.1103/PhysRevLett.99.116103}.

\bibitem{Stannard2008}
Stannard A, Martin CP, Pauliac-Vaujour E, Moriarty P, Thiele U.
\newblock Dual-Scale Pattern Formation in Nanoparticle Assemblies.
\newblock The Journal of Physical Chemistry C. 2008 09;112(39):15195--15203.
\newblock Available from: \url{https://doi.org/10.1021/jp803399d}.

\bibitem{Stannard2011}
Stannard A.
\newblock Dewetting-mediated pattern formation in nanoparticle assemblies.
\newblock Journal of Physics: Condensed Matter. 2011 feb;23(8):083001.
\newblock Available from: \url{https://doi.org/10.1088/0953-8984/23/8/083001}.

\bibitem{Chiutu2012}
Chiutu C, Sweetman AM, Lakin AJ, Stannard A, Jarvis S, Kantorovich L, et~al.
\newblock Precise Orientation of a Single ${\mathrm{C}}_{60}$ Molecule on the
  Tip of a Scanning Probe Microscope.
\newblock Phys Rev Lett. 2012 Jun;108:268302.
\newblock Available from:
  \url{https://link.aps.org/doi/10.1103/PhysRevLett.108.268302}.

\bibitem{Sweetman2016}
Sweetman A, Rashid MA, Jarvis SP, Dunn JL, Rahe P, Moriarty P.
\newblock Visualizing the orientational dependence of an intermolecular
  potential.
\newblock Nat Commun. 2016 Feb;7(1):10621.

\bibitem{London1930}
London F.
\newblock Zur Theorie und Systematik der Molekularkraefte.
\newblock Eur Phys J A. 1930 Mar;63(3-4):245--279.

\bibitem{ostwald1901}
Ostwald W.
\newblock Blocking of Ostwald ripening allowing long-term stabilization.
\newblock Phys Chem. 1901;37:385.

\bibitem{Girifalco1992}
Girifalco LA.
\newblock Molecular properties of fullerene in the gas and solid phases.
\newblock J Phys Chem. 1992 Jan;96(2):858--861.

\bibitem{Giessibl1998}
Giessibl FJ.
\newblock High-speed force sensor for force microscopy and profilometry
  utilizing a quartz tuning fork.
\newblock Applied Physics Letters. 1998 12;73(26):3956--3958.
\newblock Available from: \url{https://doi.org/10.1063/1.122948}.

\bibitem{Giessibl2011}
Giessibl FJ, Pielmeier F, Eguchi T, An T, Hasegawa Y.
\newblock Comparison of force sensors for atomic force microscopy based on
  quartz tuning forks and length-extensional resonators.
\newblock Phys Rev B. 2011 Sep;84:125409.
\newblock Available from:
  \url{https://link.aps.org/doi/10.1103/PhysRevB.84.125409}.

\bibitem{Giessibl2019}
Giessibl FJ.
\newblock The qPlus sensor, a powerful core for the atomic force microscope.
\newblock Review of Scientific Instruments. 2019 01;90(1):011101.
\newblock Available from: \url{https://doi.org/10.1063/1.5052264}.

\bibitem{qplussensor}
Giessibl FJ. The qPlus{\textregistered} Sensor.
\newblock \url{https://qplussensor.com/} 2025.
\newblock Accessed: 2026-08-21.

\bibitem{Schull2009}
Schull G, Frederiksen T, Brandbyge M, Berndt R.
\newblock Passing Current through Touching Molecules.
\newblock Phys Rev Lett. 2009 Nov;103:206803.
\newblock Available from:
  \url{https://link.aps.org/doi/10.1103/PhysRevLett.103.206803}.

\bibitem{Schull2011}
Schull G, Frederiksen T, Arnau A, S{\'a}nchez-Portal D, Berndt R.
\newblock Atomic-scale engineering of electrodes for single-molecule contacts.
\newblock Nat Nanotechnol. 2011 Jan;6(1):23--27.

\bibitem{Lakin2013}
Lakin AJ, Chiutu C, Sweetman AM, Moriarty P, Dunn JL.
\newblock Recovering molecular orientation from convoluted orbitals.
\newblock Phys Rev B. 2013 Jul;88:035447.
\newblock Available from:
  \url{https://link.aps.org/doi/10.1103/PhysRevB.88.035447}.

\bibitem{SaderJarvis}
Sader JE, Jarvis SP.
\newblock Accurate formulas for interaction force and energy in frequency
  modulation force spectroscopy.
\newblock Applied Physics Letters. 2004 03;84(10):1801--1803.
\newblock Available from: \url{https://doi.org/10.1063/1.1667267}.

\bibitem{Sweetman2014b}
Sweetman A, Stannard A.
\newblock Uncertainties in forces extracted from non-contact atomic force
  microscopy measurements by fitting of long-range background forces.
\newblock Beilstein J Nanotechnol. 2014 Apr;5:386--393.

\bibitem{Cepek1999}
Cepek C, Schiavuta P, Sancrotti M, Pedio M.
\newblock Photoemission study of ${\mathrm{C}}_{60}/\mathrm{S}\mathrm{i}(111)$
  adsorption as a function of coverage and annealing temperature.
\newblock Phys Rev B. 1999 Jul;60:2068--2073.
\newblock Available from:
  \url{https://link.aps.org/doi/10.1103/PhysRevB.60.2068}.

\bibitem{Gangopadhyay2009}
Gangopadhyay S, Woolley RAJ, Danza R, Phillips MA, Schulte K, Wang L, et~al.
\newblock C60 submonolayers on the Si(111)-(7×7) surface: Does a mixture of
  physisorbed and chemisorbed states exist?
\newblock Surface Science. 2009;603(18):2896--2901.
\newblock Available from:
  \url{https://www.sciencedirect.com/science/article/pii/S0039602809005184}.

\bibitem{Sweetman2014}
Sweetman AM, Jarvis SP, Sang H, Lekkas I, Rahe P, Wang Y, et~al.
\newblock Mapping the force field of a hydrogen-bonded assembly.
\newblock Nat Commun. 2014 May;5(1):3931.

\bibitem{Rahe2011}
Rahe P, Schütte J, Schniederberend W, Reichling M, Abe M, Sugimoto Y, et~al.
\newblock Flexible drift-compensation system for precise 3D force mapping in
  severe drift environments.
\newblock Review of Scientific Instruments. 2011 06;82(6):063704.
\newblock Available from: \url{https://doi.org/10.1063/1.3600453}.

\bibitem{Zhang2013}
Zhang J, Chen P, Yuan B, Ji W, Cheng Z, Qiu X.
\newblock Real-Space Identification of Intermolecular Bonding with Atomic Force
  Microscopy.
\newblock Science. 2013;342(6158):611--614.
\newblock Available from:
  \url{https://www.science.org/doi/abs/10.1126/science.1242603}.

\bibitem{Gross2009}
Gross L, Mohn F, Moll N, Liljeroth P, Meyer G.
\newblock The chemical structure of a molecule resolved by atomic force
  microscopy.
\newblock Science. 2009 Aug;325(5944):1110--1114.

\bibitem{Sweetman2014PRB}
Sweetman A, Jarvis SP, Rahe P, Champness NR, Kantorovich L, Moriarty P.
\newblock Intramolecular bonds resolved on a semiconductor surface.
\newblock Phys Rev B. 2014 Oct;90:165425.
\newblock Available from:
  \url{https://link.aps.org/doi/10.1103/PhysRevB.90.165425}.

\bibitem{DPGabstract}
Hapala P, Kichin G, Tautz S, Temirov R, Jelinek P.
\newblock Origin of sharp apparent intermolecular bonds in {AFM} and {STM}
  experiments.
\newblock In: DPG-Fr{\"u}hjahrstagung, Dresden; 2014. Talk O~48.8.
\newblock Available from:
  \url{https://www.dpg-verhandlungen.de/year/2014/conference/dresden/static/o48.pdf}.

\bibitem{Hapala2014}
Hapala P, Kichin G, Wagner C, Tautz FS, Temirov R, Jel\'{\i}nek P.
\newblock Mechanism of high-resolution STM/AFM imaging with functionalized
  tips.
\newblock Phys Rev B. 2014 Aug;90:085421.
\newblock Available from:
  \url{https://link.aps.org/doi/10.1103/PhysRevB.90.085421}.

\bibitem{Oinonen2024}
Oinonen N, Yakutovich AV, Gallardo A, Ondr{\'a}{\v c}ek M, Hapala P, Krej{\v
  c}{\'\i} O.
\newblock Advancing scanning probe microscopy simulations: A decade of
  development in probe-particle models.
\newblock Comput Phys Commun. 2024 Dec;305(109341):109341.

\bibitem{Hamalainen2014}
H\"am\"al\"ainen SK, van~der Heijden N, van~der Lit J, den Hartog S, Liljeroth
  P, Swart I.
\newblock Intermolecular Contrast in Atomic Force Microscopy Images without
  Intermolecular Bonds.
\newblock Phys Rev Lett. 2014 Oct;113:186102.
\newblock Available from:
  \url{https://link.aps.org/doi/10.1103/PhysRevLett.113.186102}.

\bibitem{Jarvis2015}
Jarvis SP, Rashid MA, Sweetman A, Leaf J, Taylor S, Moriarty P, et~al.
\newblock Intermolecular artifacts in probe microscope images of
  ${\mathrm{C}}_{60}$ assemblies.
\newblock Phys Rev B. 2015 Dec;92:241405(R).
\newblock Available from:
  \url{https://link.aps.org/doi/10.1103/PhysRevB.92.241405}.

\bibitem{Jarvis2015review}
Jarvis SP.
\newblock Resolving intra- and inter-molecular structure with non-contact
  atomic force microscopy.
\newblock Int J Mol Sci. 2015 Aug;16(8):19936--19959.

\bibitem{Adams2020BBC}
{BBC Radio 4}. 42 {Douglas Adams} Quotes to Live By.
\newblock
  \url{https://www.bbc.co.uk/programmes/articles/2bcFfMt6rGLTPpbG0yLwPw0/42-douglas-adams-quotes-to-live-by}
  2020.
\newblock Radio 4 in Four.

\bibitem{Moll2012}
Moll N, Gross L, Mohn F, Curioni A, Meyer G.
\newblock A simple model of molecular imaging with noncontact atomic force
  microscopy.
\newblock New J Phys. 2012 Aug;14(8):083023.

\bibitem{Jarvis2014Pauli}
Jarvis S, Sweetman A, Kantorovich L, McGlynn E, Moriarty P. Pauli's Principle
  in Probe Microscopy 2014.
\newblock Available from: \url{https://arxiv.org/abs/1408.1026}.

\bibitem{Brand2019}
Brand J, N{\'e}el N, Kr{\"o}ger J.
\newblock Probing relaxations of atomic-scale junctions in the Pauli repulsion
  range.
\newblock New J Phys. 2019 Oct;21(10):103041.

\bibitem{Verwoerd1980}
Verwoerd WS.
\newblock Cluster calculations of the surface dimer structure on Si(100)
  surfaces.
\newblock Surface Science. 1980;99(3):581--597.
\newblock Available from:
  \url{https://www.sciencedirect.com/science/article/pii/0039602880905555}.

\bibitem{Robles2012}
Robles R, Kepenekian M, Monturet S, Joachim C, Lorente N.
\newblock Energetics and stability of dangling-bond silicon wires on H
  passivated Si(100).
\newblock Journal of Physics: Condensed Matter. 2012 sep;24(44):445004.
\newblock Available from: \url{https://doi.org/10.1088/0953-8984/24/44/445004}.

\bibitem{Sweetman2011}
Sweetman A, Jarvis S, Danza R, Bamidele J, Gangopadhyay S, Shaw GA, et~al.
\newblock Toggling Bistable Atoms via Mechanical Switching of Bond Angle.
\newblock Phys Rev Lett. 2011 Mar;106:136101.
\newblock Available from:
  \url{https://link.aps.org/doi/10.1103/PhysRevLett.106.136101}.

\bibitem{Sweetman2011PRB}
Sweetman A, Jarvis S, Danza R, Bamidele J, Kantorovich L, Moriarty P.
\newblock Manipulating Si(100) at 5 K using qPlus frequency modulated atomic
  force microscopy: Role of defects and dynamics in the mechanical switching of
  atoms.
\newblock Phys Rev B. 2011 Aug;84:085426.
\newblock Available from:
  \url{https://link.aps.org/doi/10.1103/PhysRevB.84.085426}.

\bibitem{JarvisPRB2012}
Jarvis S, Sweetman A, Bamidele J, Kantorovich L, Moriarty P.
\newblock Role of orbital overlap in atomic manipulation.
\newblock Phys Rev B. 2012 Jun;85:235305.
\newblock Available from:
  \url{https://link.aps.org/doi/10.1103/PhysRevB.85.235305}.

\bibitem{Jarvis2013Beilstein}
Jarvis SP, Kantorovich L, Moriarty P.
\newblock Structural development and energy dissipation in simulated silicon
  apices.
\newblock Beilstein J Nanotechnol. 2013 Dec;4:941--948.

\bibitem{Sweetman2020}
Sweetman A, Harivyasi SS, Henry J, Lekkas I.
\newblock Imaging and manipulation of adsorbed Pb adatom structures on the
  Si(100) surface by noncontact atomic force microscopy.
\newblock Phys Rev B. 2020 Dec;102:235417.
\newblock Available from:
  \url{https://link.aps.org/doi/10.1103/PhysRevB.102.235417}.

\bibitem{Sugimoto2008}
Sugimoto Y, Pou P, Custance O, Jelinek P, Abe M, Perez R, et~al.
\newblock Complex patterning by vertical interchange atom manipulation using
  atomic force microscopy.
\newblock Science. 2008 Oct;322(5900):413--417.

\bibitem{EiglerSwitch}
Eigler DM, Lutz CP, Rudge WE.
\newblock An atomic switch realized with the scanning tunnelling microscope.
\newblock Nature. 1991 Aug;352(6336):600--603.

\bibitem{Becker1987}
Becker RS, Golovchenko JA, Swartzentruber BS.
\newblock Atomic-scale surface modifications using a tunnelling microscope.
\newblock Nature. 1987 Jan;325(6103):419--421.

\bibitem{Lyo1991}
Lyo IW, Avouris P.
\newblock Field-induced nanometer- to atomic-scale manipulation of silicon
  surfaces with the {STM}.
\newblock Science. 1991 Jul;253(5016):173--176.

\bibitem{Shen1995}
Shen TC, Wang C, Abeln GC, Tucker JR, Lyding JW, Avouris P, et~al.
\newblock Atomic-scale desorption through electronic and vibrational excitation
  mechanisms.
\newblock Science. 1995 Jun;268(5217):1590--1592.

\bibitem{Pavlicek2017-hh}
Pavli{\v c}ek N, Majzik Z, Meyer G, Gross L.
\newblock Tip-induced passivation of dangling bonds on hydrogenated Si(100)-2
  $\times$ 1.
\newblock Appl Phys Lett. 2017 Jul;111(5):053104.

\bibitem{Huff2017}
Huff TR, Labidi H, Rashidi M, Koleini M, Achal R, Salomons MH, et~al.
\newblock Atomic white-out: Enabling atomic circuitry through mechanically
  induced bonding of single hydrogen atoms to a silicon surface.
\newblock ACS Nano. 2017 Sep;11(9):8636--8642.

\bibitem{Sharp2012}
Sharp P, Jarvis S, Woolley R, Sweetman A, Kantorovich L, Pakes C, et~al.
\newblock Identifying passivated dynamic force microscopy tips on {H:Si(100}).
\newblock Appl Phys Lett. 2012 Jun;100(23):233120.

\bibitem{Eigler1990}
Eigler DM, {E  K  Schweizer}.
\newblock Positioning single atoms with a scanning tunnelling microscope.
\newblock Nature. 1990 Apr;344(6266):524--526.

\bibitem{Crommie1993}
Crommie MF, Lutz CP, Eigler DM.
\newblock Confinement of electrons to quantum corrals on a metal surface.
\newblock Science. 1993 Oct;262(5131):218--220.

\bibitem{Manoharan2000}
Manoharan HC, Lutz CP, Eigler DM.
\newblock Quantum mirages formed by coherent projection of electronic
  structure.
\newblock Nature. 2000 Feb;403(6769):512--515.

\bibitem{Khajetoorians2019}
Khajetoorians AA, Wegner D, Otte AF, Swart I.
\newblock Creating designer quantum states of matter atom-by-atom.
\newblock Nat Rev Phys. 2019 Sep;1(12):703--715.

\bibitem{Sierda2023}
Sierda E, Huang X, Badrtdinov DI, Kiraly B, Knol EJ, Groenenboom GC, et~al.
\newblock Quantum simulator to emulate lower-dimensional molecular structure.
\newblock Science. 2023 Jun;380(6649):1048--1052.

\bibitem{AbbasiPerez2021}
Abbasi-P{\'e}rez D, Sang H, Junqueira FLQ, Sweetman A, Recio JM, Moriarty P,
  et~al.
\newblock Cyclic single atom vertical manipulation on a nonmetallic surface.
\newblock J Phys Chem Lett. 2021 Nov;12(46):11383--11390.

\bibitem{Custance2009}
Custance O, Perez R, Morita S.
\newblock Atomic force microscopy as a tool for atom manipulation.
\newblock Nat Nanotechnol. 2009 Dec;4(12):803--810.

\bibitem{Folsch1}
F\"olsch S, Hyldgaard P, Koch R, Ploog KH.
\newblock Quantum Confinement in Monatomic Cu Chains on Cu(111).
\newblock Phys Rev Lett. 2004 Feb;92:056803.
\newblock Available from:
  \url{https://link.aps.org/doi/10.1103/PhysRevLett.92.056803}.

\bibitem{Folsch2}
Lagoute J, Liu X, F\"olsch S.
\newblock Electronic properties of straight, kinked, and branched
  $\mathrm{Cu}∕\mathrm{Cu}(111)$ quantum wires: A low-temperature scanning
  tunneling microscopy and spectroscopy study.
\newblock Phys Rev B. 2006 Sep;74:125410.
\newblock Available from:
  \url{https://link.aps.org/doi/10.1103/PhysRevB.74.125410}.

\bibitem{Folsch3}
D{\'\i}az-Tendero S, F{\"o}lsch S, Olsson FE, Borisov AG, Gauyacq JP.
\newblock Electron propagation along Cu nanowires supported on a Cu(111)
  surface.
\newblock Nano Lett. 2008 Sep;8(9):2712--2717.

\bibitem{Rubio1996}
Rubio G, Agra\"{\i}t N, Vieira S.
\newblock Atomic-Sized Metallic Contacts: Mechanical Properties and Electronic
  Transport.
\newblock Phys Rev Lett. 1996 Mar;76:2302--2305.
\newblock Available from:
  \url{https://link.aps.org/doi/10.1103/PhysRevLett.76.2302}.

\bibitem{Ternes2011}
Ternes M, Gonz\'alez C, Lutz CP, Hapala P, Giessibl FJ, Jel\'{\i}nek P, et~al.
\newblock Interplay of Conductance, Force, and Structural Change in Metallic
  Point Contacts.
\newblock Phys Rev Lett. 2011 Jan;106:016802.
\newblock Available from:
  \url{https://link.aps.org/doi/10.1103/PhysRevLett.106.016802}.

\bibitem{Hauptmann2012}
Hauptmann N, Mohn F, Gross L, Meyer G, Frederiksen T, Berndt R.
\newblock Force and conductance during contact formation to a C60molecule.
\newblock New J Phys. 2012 Jul;14(7):073032.

\bibitem{Sperl2011}
Sperl A, Kr{\"o}ger J, Berndt R.
\newblock Controlled metalation of a single adsorbed phthalocyanine.
\newblock Angew Chem Int Ed Engl. 2011 May;50(23):5294--5297.

\bibitem{Sperl2011JACS}
Sperl A, Kr{\"o}ger J, Berndt R.
\newblock Demetalation of a single organometallic complex.
\newblock J Am Chem Soc. 2011 Jul;133(29):11007--11009.

\bibitem{SawReview}
Hla SW.
\newblock Scanning tunneling microscopy single atom/molecule manipulation and
  its application to nanoscience and technology.
\newblock J Vac Sci Technol B Microelectron Nanometer Struct Process Meas
  Phenom. 2005 Jul;23(4):1351--1360.

\bibitem{NanosurfLab}
{Nanosurf Lab}. Gallery.
\newblock \url{https://nanosurf.fzu.cz/images.php}Institute of Physics, Czech
  Academy of Sciences (FZU).

\bibitem{Bartels1997}
Bartels L, Meyer G, Rieder KH.
\newblock Controlled vertical manipulation of single {CO} molecules with the
  scanning tunneling microscope: A route to chemical contrast.
\newblock Appl Phys Lett. 1997 Jul;71(2):213--215.

\bibitem{Emmrich2015}
Emmrich M, Huber F, Pielmeier F, Welker J, Hofmann T, Schneiderbauer M, et~al.
\newblock Surface structure. Subatomic resolution force microscopy reveals
  internal structure and adsorption sites of small iron clusters.
\newblock Science. 2015 Apr;348(6232):308--311.

\bibitem{Repp2003}
Repp J, Meyer G, Rieder KH, Hyldgaard P.
\newblock Site Determination and Thermally Assisted Tunneling in Homogenous
  Nucleation.
\newblock Phys Rev Lett. 2003 Nov;91:206102.
\newblock Available from:
  \url{https://link.aps.org/doi/10.1103/PhysRevLett.91.206102}.

\bibitem{GerberKavli}
Gerber C. Not Even the Sky is the Limit.
\newblock \url{https://www.kavliprize.org/christoph-gerber-autobiography} 2016.
\newblock Autobiographical essay, The Kavli Prize.

\bibitem{Giessibl2000}
Giessibl FJ, Hembacher S, Bielefeldt H, Mannhart J.
\newblock Subatomic features on the silicon (111)-(7x7) surface observed by
  atomic force microscopy.
\newblock Science. 2000 Jul;289(5478):422--426.

\bibitem{Moller2026}
Møller M, Rahe P, Ghaderzadeh S, Besley E, Moriarty P. Static heterogeneity
  generates apparent universality in first-passage bursty dynamics 2026.
\newblock Available from: \url{https://arxiv.org/abs/2604.15084}.

\bibitem{Barabasi2005}
Barab{\'a}si AL.
\newblock The origin of bursts and heavy tails in human dynamics.
\newblock Nature. 2005 May;435(7039):207--211.

\bibitem{Carlisle2026NCAFM}
Carlisle F, Seeja~Sivakumar N, Phillips O, James T, Civita D, Sufyan A, et~al.
\newblock Invasive and non-invasive probes of sub-{K} molecular diffusion.
\newblock In: NC-AFM 2026: 27th International Conference on Non-Contact Atomic
  Force Microscopy. Innsbruck, Austria; 2026. Conference abstract.
\newblock Available from:
  \url{https://www.uibk.ac.at/media/filer_public/59/f7/59f7531d-7534-400c-9ee0-ce1368efb819/ncafm2026___fullprogram-compressed.pdf}.

\bibitem{Donato1}
Civita D, Kolmer M, Simpson GJ, Li AP, Hecht S, Grill L.
\newblock Control of long-distance motion of single molecules on a surface.
\newblock Science. 2020 Nov;370(6519):957--960.

\bibitem{Bader2000}
Bader RFW.
\newblock Atomic force microscope as an open system and the Ehrenfest force.
\newblock Phys Rev B. 2000 Mar;61:7795--7802.
\newblock Available from:
  \url{https://link.aps.org/doi/10.1103/PhysRevB.61.7795}.

\bibitem{Drexler1992}
Drexler KE.
\newblock Nanosystems: Molecular Machinery, Manufacturing, and Computation.
\newblock New York: Wiley-Interscience; 1992.

\bibitem{Kalinin2015}
Kalinin SV, Sumpter BG, Archibald RK.
\newblock Big-deep-smart data in imaging for guiding materials design.
\newblock Nat Mater. 2015 Oct;14(10):973--980.

\bibitem{Rickman2019}
Rickman JM, Lookman T, Kalinin SV.
\newblock Materials informatics: From the atomic-level to the continuum.
\newblock Acta Mater. 2019 Apr;168:473--510.

\bibitem{Kalinin2022}
Kalinin SV, Ophus C, Voyles PM, Erni R, Kepaptsoglou D, Grillo V, et~al.
\newblock Machine learning in scanning transmission electron microscopy.
\newblock Nat Rev Methods Primers. 2022 Mar;2(1).

\bibitem{Madika2025}
Madika B, Saha A, Kang C, Buyantogtokh B, Agar J, Wolverton CM, et~al.
\newblock Artificial intelligence for materials discovery, development, and
  optimization.
\newblock ACS Nano. 2025 Aug;19(30):27116--27158.

\bibitem{Gordon2019}
Gordon O, D'Hondt P, Knijff L, Freeney SE, Junqueira F, Moriarty P, et~al.
\newblock Scanning tunneling state recognition with multi-class neural network
  ensembles.
\newblock Rev Sci Instrum. 2019 Oct;90(10):103704.

\bibitem{Gordon2020a}
Gordon OM, Junqueira FLQ, Moriarty PJ.
\newblock Embedding human heuristics in machine-learning-enabled probe
  microscopy.
\newblock Mach Learn Sci Technol. 2020 Mar;1(1):015001.

\bibitem{Gordon2020b}
Gordon OM, Hodgkinson JEA, Farley SM, Hunsicker EL, Moriarty PJ.
\newblock Automated searching and identification of self-organized
  nanostructures.
\newblock Nano Lett. 2020 Oct;20(10):7688--7693.

\bibitem{Farley2020}
Farley S, Hodgkinson JEA, Gordon OM, Turner J, Soltoggio A, Moriarty PJ, et~al.
\newblock Improving the segmentation of scanning probe microscope images using
  convolutional neural networks.
\newblock Mach Learn Sci Technol. 2020 Dec;2(1):015015.

\bibitem{Celotta2014}
Celotta RJ, Balakirsky SB, Fein AP, Hess FM, Rutter GM, Stroscio JA.
\newblock Invited Article: Autonomous assembly of atomically perfect
  nanostructures using a scanning tunneling microscope.
\newblock Rev Sci Instrum. 2014 Dec;85(12):121301.

\bibitem{Kalff2016}
Kalff FE, Rebergen MP, Fahrenfort E, Girovsky J, Toskovic R, Lado JL, et~al.
\newblock A kilobyte rewritable atomic memory.
\newblock Nat Nanotechnol. 2016 Nov;11(11):926--929.

\bibitem{Leinen2020}
Leinen P, Esders M, Sch{\"u}tt KT, Wagner C, M{\"u}ller KR, Tautz FS.
\newblock Autonomous robotic nanofabrication with reinforcement learning.
\newblock Sci Adv. 2020 Sep;6(36):eabb6987.

\bibitem{Ramsauer2023}
Ramsauer B, Simpson GJ, Cartus JJ, Jeindl A, Garc{\'\i}a-L{\'o}pez V, Tour JM,
  et~al.
\newblock Autonomous single-molecule manipulation based on reinforcement
  learning.
\newblock J Phys Chem A. 2023 Mar;127(8):2041--2050.

\bibitem{Chen2022}
Chen IJ, Aapro M, Kipnis A, Ilin A, Liljeroth P, Foster AS.
\newblock Precise atom manipulation through deep reinforcement learning.
\newblock Nat Commun. 2022 Dec;13(1):7499.

\bibitem{Woolley2011}
Woolley RAJ, Stirling J, Radocea A, Krasnogor N, Moriarty P.
\newblock Automated probe microscopy via evolutionary optimization at the
  atomic scale.
\newblock Appl Phys Lett. 2011 Jun;98(25):253104.

\bibitem{Alldritt2022}
Alldritt B, Urtev F, Oinonen N, Aapro M, Kannala J, Liljeroth P, et~al.
\newblock Automated tip functionalization via machine learning in scanning
  probe microscopy.
\newblock Comput Phys Commun. 2022 Apr;273(108258):108258.

\bibitem{Rashidi2018}
Rashidi M, Wolkow RA.
\newblock Autonomous scanning probe microscopy in situ tip conditioning through
  machine learning.
\newblock ACS Nano. 2018 Jun;12(6):5185--5189.

\bibitem{Wang2021}
Wang S, Zhu J, Blackwell R, Fischer FR.
\newblock Automated tip conditioning for scanning tunneling spectroscopy.
\newblock J Phys Chem A. 2021 Feb;125(6):1384--1390.

\bibitem{Ternes2008}
Ternes M, Lutz CP, Hirjibehedin CF, Giessibl FJ, Heinrich AJ.
\newblock The Force Needed to Move an Atom on a Surface.
\newblock Science. 2008;319(5866):1066--1069.
\newblock Available from:
  \url{https://www.science.org/doi/abs/10.1126/science.1150288}.

\bibitem{BinnigBook}
Binnig G.
\newblock Aus dem Nichts: {\"U}ber die Kreativit{\"a}t von Natur und Mensch.
\newblock M{\"u}nchen: Piper; 1989.
\newblock With drawings and poems by Rudi Gerharz.

\end{thebibliography}

\end{document}